\documentclass{aa}  

\usepackage{graphicx}
\usepackage{txfonts}
\usepackage{graphicx}	% Including figure files
\usepackage{amsmath}	% Advanced maths commands
\usepackage{amssymb}	% Extra maths symbols
\usepackage{color}
\usepackage{soul}
\usepackage{natbib}
\usepackage{booktabs}
\usepackage{subfig}
\usepackage{hyperref}
\usepackage{float}
\restylefloat{table}
\usepackage{placeins}

\begin{document}

   \title{A %comprehensive 
   population synthesis study of the {\it Gaia} 100 pc unresolved white dwarf-main sequence binary population}

   \author{A. Santos-García\inst{1},
          S. Torres\inst{1,2}\thanks{E-mail: santiago.torres@upc.edu},
          A. Rebassa-Mansergas\inst{1,2},
          A. J. Brown\inst{1}
          }
\titlerunning{{\it Gaia} 100 pc unresolved white dwarf- main sequence binary population}
\authorrunning{Santos-García et al.}

   \institute{Departament de F\'{\i}sica, Universitat Polit\`{e}cnica de Catalunya, c/ Esteve Terrades 5, 08860 Castelldefels, Barcelona, Spain\\        
         \and
             Institut d'Estudis Espacials de Catalunya, Esteve Terradas, 1, Edifici RDIT, Campus PMT-UPC, 08860 Castelldefels, Barcelona, Spain\\
            }

   \date{Accepted XXX. Received YYY; in original form ZZZ}

% \abstract{}{}{}{}{} 
% 5 {} token are mandatory
 
  \abstract
  {Binary stars consisting of a white dwarf and a main sequence star, WDMS, are excellent tools to study a wide variety of open problems in modern astronomy. However, due to selection effects, the currently known population of such objects is severely affected by observational biases. This is particularly the case for unresolved systems in which the main sequence companions usually outshine the white dwarf components.}
   {This work aims to comprehensively simulate the population of unresolved WDMS binaries within 100 pc of the Sun and to compare the outcome with the currently most complete volume-limited sample available from {\it Gaia} data. By doing so, we seek to refine our understanding of WDMS formation and evolution and to test the theoretical models against the observed data. }
   {We employ a population synthesis code, \texttt{MRBIN}, extensively developed by our group and based on Monte Carlo techniques, which uses a standard binary stellar evolutionary code adapted to cover a wide range of stars across all ages, masses, and metallicities. Different physical processes such as mass transfer, common envelope evolution, and tidal interactions are considered. Selection criteria matching those of {\it Gaia} observations are applied to generate synthetic populations comparable to the observed WDMS sample.} 
   {Our analysis provides overall fractions of single main sequence stars and white dwarfs, and resolved/unresolved WDMS in excellent agreement with observed values. The synthetic data accurately populate the expected regions in the {\it Gaia} color-magnitude diagram. However,  simulations predict a lower number of extremely low-mass white dwarfs, suggesting potential issues in observed mass derivations. Additionally, our analysis constrains the common envelope efficiency to 0.1–0.4, consistent with previous findings, and estimates a total completeness of about 25\% for the observed sample, confirming the strong observational limitations for unresolved WDMS.}
   {This work provides understanding into WDMS binary evolution and highlights limitations in observational detectability, underscoring the importance of fine-tuning parameters in binary evolution models to improve population synthesis studies.}


   \maketitle
%
%________________________________________________________________

%%%%%%%%%%%%%%%%%%%%%%%%%%%%%%%%%%%%%%%%%%%%%%%%%%

%%%%%%%%%%%%%%%%% BODY OF PAPER %%%%%%%%%%%%%%%%%%

\section{Introduction}
\label{s:intro}

More than 97\% of all main sequence stars in our Galaxy (those with M$\leq$ 8--11 M$_{\odot}$) will become or have already become white dwarfs. With the cessation of nuclear reactions in their interiors, these stellar remnants slowly cool down while releasing their thermal energy. This cooling process can extend over more than 10 billion years, which is why white dwarfs provide important information about the history of our Galaxy and its constituents \citep{Althaus+10, GarciaBerro+Oswalt+16, Isern+22}.

White dwarfs are usually found isolated, but they can also be part of binary systems. In progenitor main-sequence binaries, each component evolves as a single star if the system is sufficiently separated \citep[$\ge10\,$AU;][]{Willems+Kolb04, Farihi+10}. The more massive star leaves the main sequence first and eventually losses mass to produce a white dwarf, thus resulting in the expansion of the orbit and the formation of a widely-separated white dwarf plus main sequence binary (WDMS). Wide WDMS binaries are excellent tools to tackle, for instance, the age-metallicity relation \citep{Rebassa2021b} and age-velocity dispersion relation \citep{Raddi+22} of the Galactic disc, and the age-activity-rotation relation of low-mass main sequence stars \citep{Rebassa2023, Chiti+24}, since white dwarf ages can be accurately derived in these systems \citep{Toloza+23}.

If, conversely, the original main sequence stars are close enough, mass transfer interactions are likely to ensue when the more massive star becomes a giant or an asymptotic giant star.  
This generally results in an unstable process of mass transfer, which leads to the formation of a common envelope surrounding both stars \citep[][]{Pac76, Webbink08}. Due to friction involving both the giant's core (i.e. the future white dwarf) and the main sequence companion with the material of the envelope, the orbital separation reduces dramatically and the orbital energy released is used to eject the envelope. As a consequence, the newly-formed WDMS is in a much closer orbit with periods ranging from a few hours to several days \citep{Nebot+11, Rebassa+12}. These post-common envelope binaries experience further angular momentum losses due to gravitational wave radiation and/or magnetic braking, which continues to shorten their orbital periods until a second phase of (stable) mass transfer begins producing exotic objects such as cataclysmic variables or super-soft X-ray sources. It is also possible that a second common envelope begins if the secondary star has time to leave the main sequence, which leads to the formation of a double white dwarf (DWD) in a tight orbit.

Close WDMS binaries are excellent tools to constrain a wide range of open problems in astronomy. To begin with, the reconstruction of their past evolution places direct constraints on how efficiently the release of orbital energy is used to expel the common envelope. Thus, since the pioneer work of \citet{Zorotovic+10}, it has been constantly found that this process is rather inefficient \citep{Camacho+14, Cojocaru+17, Toonen+13, Iaconi+19, Zorotovic+22}. By predicting the future evolution of close WDMS, it is also possible to analyze their different evolutionary paths that, ultimately, can be used to place constraints of the different channels towards type Ia supernovae \citep{Parsons+16, Parsons+2023, Hernandez+21, Hernandez+22}. Moreover, eclipsing WDMS allow deriving independent masses and radii for both components, which permits testing the theoretical mass-radius relation of white dwarfs \citep{Parsons+17, Brown+23}, main sequence stars \citep{Parsons+18} and even sub-dwarf stars \citep{Rebassa+19} and brown dwarfs \citep{French+24}.

Despite the significance of WDMS it is important to bear in mind that the currently known population is severely affected by selection effects. This is particularly the case for unresolved systems since the main sequence components generally outshine the white dwarfs in the optical \citep{Rebassa+10}. One way to overcome this issue is by making use of ultraviolet spectroscopy and/or photometry, which minimizes the contribution of the companions at those wavelengths \citep{Parsons+16, Rebassa+17, Ren+20, Anguiano+22, Nayak+24, Sidharth+24}. An additional problem is that most WDMS catalogs are the result of mining data from magnitude-limited samples and, to make things worse, in most cases the WDMS were observed following selection criteria designed for targeting different objects. Volume-limited samples have been possible to achieve thanks to the excellent photometry and astrometry provided by \emph{Gaia} \citep{Inight+21}. Using the early data release 3, \citet{Rebassa+21} acquired and characterized the most homogeneous volume-limited sample of unresolved WDMS binaries to date, within 100\,pc. Although this sample is still far from complete -- since it only includes the bridge region in the Gaia color-magnitude diagram between single main-sequence stars and single white dwarfs -- it constitutes a good benchmark for testing theoretical models and assumptions by generating synthetic populations for comparison.

In this sense, binary population synthesis studies have been widely used, particularly in the study of DWD and WDMS binary systems \citep[e.g.][]{Nelemans2001a,Nelemans2001b,Toonen+2014,Toonen+2017,Camacho+14,Cojocaru+17,Canals+18}. More recently, \citet{Torres+22}, taking advantage of the accurate astrometric and photometric data provided by the {\it Gaia} mission, succeeded in tuning the model parameters so that the synthetic population reproduces the observed percentages of resolved double-degenerate and WDMS binary systems. Among these parameters, they identified accurate prescriptions for the binary fraction, initial mass ratio distribution, and initial separation distribution, among others.

The population synthesis code used in \citet{Torres+22}, largely developed by our group, incorporates a comprehensive description of the most relevant physical processes in stellar binary evolution (mass transfer, tidal interactions, stellar wind losses, magnetic braking, gravitational radiation, etc.), as well as a realistic characterization of the different Galactic components --thin and thick disks, and halo-- for a wide range of masses and metallicities. In addition, a detailed set of selection criteria has been implemented to achieve significant statistical comparisons between the observed and synthetic samples.

In this paper, we aim to use this consolidated binary population synthesis code \citep{Torres+22} to study the 100 pc unresolved WDMS sample derived from {\it Gaia} data \citep{Rebassa2021b}. In particular, we intend to evaluate the degree of completeness of the observed sample and to derive some insights into the evolution of close binary systems, especially those parameters related to the common-envelope phase.

In Section\,\ref{s:gaia} we describe the \emph{Gaia} WDMS binary sample. In Section\,\ref{s:model} we provide details of our Monte Carlo simulator, which produces synthetic samples of WDMS to be directly compared to the observed one. In Section\,\ref{s:pop} we present our results and we conclude in Section\,\ref{s:concl}.

\section{The {\it Gaia} 100 pc WDMS binary population}
\label{s:gaia}

To test the models used and the assumptions made in our simulations (details are provided in Section\,\ref{s:model}), we aim to compare the synthetic sample of unresolved WDMS binaries with the observed population obtained from \emph{Gaia} within 100 pc. This distance limit is imposed with the aim of dealing with the largest nearly-complete observed volume-limited sample \citep{Jimenez+18}. To that end, we adopt the unresolved WDMS binary catalog of \citet{Rebassa+21}, who searched for candidates within the \emph{Gaia} color-magnitude region  between the single white dwarf and the single main sequence star loci (this is represented by solid blue lines in Figure\,\ref{f:Gaiavssim}). The sample initially included all \emph{Gaia} sources with parallax and the three band relative error fluxes under 10\%. Moreover, objects with a large \emph{excess factor} were excluded from the list following \citet{Riello+21}, which resulted in 2001 WDMS binary candidates within the considered region. To further exclude contaminants, \citet{Rebassa+21} used VOSA (Virtual Observatory SED Analyser; \citealt{Bayo+08}) to fit the SEDs of the 2001 candidates with single white dwarf models, single main sequence star models and a combination of both. Hence,  only those with two-body fits and with VOSA parameter ${\rm Vgf_b}$\footnote{Visual goodness of fit: \url{http://svo2.cab.inta-csic.es/theory/vosa/helpw4.php?otype=star&action=help&what=fit}}$<15$  were kept. During this exercise, \citet{Rebassa+21} considered only objects with enough measurements to carry out the SED fitting and without wrong photomeric values. This reduced the sample to 280 targets, of which 129 were further excluded due to bad fitting arising as a consequence of ultraviolet excess likely arising from magnetic activity of single main sequence stars (3), bad SED fitting (90) or contamination from nearby sources (36). Closer inspection of the remaining 151 candidates revealed that 34 had unreliable \emph{Gaia} astrometric solutions (RUWE >2, or astrometric\_excess\_noise >2 and astrometric\_excess\_noise\_sig >2) and were also excluded. Finally, 5 objects turned out to be cataclysmic variables thus leaving 112 good WDMS binary candidates.

The VOSA two-body fits yielded the following stellar parameters for both components of these 112 objects: effective temperature, radius and bolometric luminosity. White dwarf masses (hence surface gravities) were obtained interpolating the effective temperatures and radii in the La Plata cooling sequences \citep[e.g.,][]{Camisassa+16,Camisassa+19}. For 98 of the 112 objects, the measured parameters were considered as reliable by \citet{Rebassa+21}.

\section{The population synthesis modeling}
\label{s:model}

\subsection{The Monte Carlo simulator}
\label{ss:MonteCarlo}

In order to generate a synthetic population of both single and binary stars in the Galaxy, we use the Monte Carlo simulator developed by our group \citep{Torres+98,GarciaBerro+99}. The code, named \texttt{MRBIN}, not only is capable of reproducing realistic single main sequence and single white dwarf Galactic populations \citep[e.g.][]{Torres+01, Torres+19}, but also allows to reproduce main sequence binary stars and their subsequent evolution \citep[e.g][]{Torres+22}. The binary stellar evolution is modeled following the binary stellar evolution code (BSE) developed by \cite{Hurley+02}, with particular updates on white dwarf binaries \citep[e.g.][]{Camacho+14, Cojocaru+17, Canals+18}.

The modeling of the entire single and binary population requires a series of input parameters that we adopted from \citet{Torres+22}. In that work, we used the same code, \texttt{MRBIN}, to perform a population synthesis fit of the current populations of resolved WDMS and DWD systems within 100\,pc that satisfactory reproduced the observed rates by \emph{Gaia}, namely 6.31\% for WDMS binary systems, and 1.18\% for DWDs. This was achieved by considering a main sequence binary fraction of 32\%; an initial mass function from 
\citet{Kroupa01} for the range  0.08-50 M$_\odot$, and extending that distribution with a positive slope according to \citet{Sollima19}, to reproduce the primary mass distribution -- the more massive star in the main sequence binary $M_{1}$--; an initial thermal eccentricity distribution following \citealp[]{Heggie75}; an orbital separation of the form $f(a)=a^{-1}$,
and a mass ratio distribution following $n(q) \sim q^{-1.13}$ (with $q=M_{1}/M_{2}$, where $M_{2}$ is the secondary, less massive, star mass). The rest of parameters can be found in \citealp[]{Torres+22}.

Moreover, the simulator randomly assigns each single and binary star to a Galactic component according to the ratio 85:10:5 for the thin disk, thick disk and halo, respectively, following  \citet{Torres+19}. Objects from the different Galactic components are modeled following different criteria regarding their ages, metallicities, spatial distributions and kinematics. In what follows, we summarize the rest of Galactic inputs, while a detailed description and their corresponding references can be found in the aforementioned paper. For the thin disk stars we adopt an exponential spatial distribution with a scale height of 250\,pc and a scale length of 2.6\,kpc, and constant star formation rate with a disk age of 9.2\,Gyr. For thick disk stars we consider a burst of star formation 10\,Gyr ago extended up to 12\,Gyr with a scale height and scale length of 1.5\,kpc and 3.5\,kpc, respectively. For halo stars we assume an isothermal spatial distribution with a constant burst of 1 Gyr that took place 12.5\,Gyr in the past. In the three cases the age-metallicity relation follows a constant model \citep{Rebassa2021b}. That is, we randomly generate metallicity values following a Gaussian distribution with a mean and a dispersion of $Z$=0.02 and 0.005 for the thin disk, $Z$=0.01 and 0.003 for the thick disk, and $Z$=0.0015 and 0.001 for the halo, respectively.

Wide binary systems evolve like single stars avoiding mass transfer episodes. However, for the evolution of binary systems with shorter separations, our code, based on the \texttt{BSE} code, uses a wide set of parameters to model the mass transfer and angular momentum losses.
In particular, we adopt a common envelope treatment characterized by the $\alpha_{\rm CE}$ parameter (\citealp{Tout+97}, \citealp{Webbink08}) during a common envelope phase, which quantifies the efficiency with which orbital energy is transferred from the binary system to the envelope of the donor star to eventually eject it. In our reference model, we use a value of $\alpha_{\rm CE}=0.3$ \citep{Camacho+14}.

Once the systems (both single and binary stars) are evolved to present time, the \texttt{MRBIN} code allows the derivation of magnitudes in a given photometric system. For the purpose of this paper (which requires a direct comparison to the observed \emph{Gaia} data; see Section\,\ref{s:gaia}), we derive the {\it Gaia} magnitudes following a three spline interpolation of luminosity, effective temperature and metallicity in the PARSEC \citep[][]{Bressan+12} evolutionary tracks
% provided by MIST \citep[][]{Dotter16} or 
for the main sequence stars. In particular, for luminosities higher than 0.1\,L$_{\odot}$, we use the effective temperature, $T_{\rm eff}$, directly provided by the \texttt{BSE} code. However, for luminosities below that value, we applied an improved correction to $T_{\rm eff}$ by performing a multi-variable interpolation using luminosity and metallicity values to obtain the corresponding $T_{\rm eff}$ according to the PARSEC evolutionary models. In any case, to avoid issues with photometric uncertainties in the models, our final simulated sample is restricted to absolute magnitudes 
$M_G<14.2\,$mag. For those systems that become white dwarfs, their cooling time is accurately determined using the most up-to-date evolutionary cooling sequences provided by the La Plata group \citep[e.g.,][]{Camisassa+16,Camisassa+19}, which take into account the different core and chemical compositions for the whole range of possible masses. The \emph{Gaia} magnitudes are interpolated within the updated grid of white dwarf atmospheric models by \citet{Koester+10}, assuming all white dwarfs to be hydrogen-rich. 

\begin{figure*}
\centering
\includegraphics[trim=0 0 0 0 clip=true, width=0.96\columnwidth]{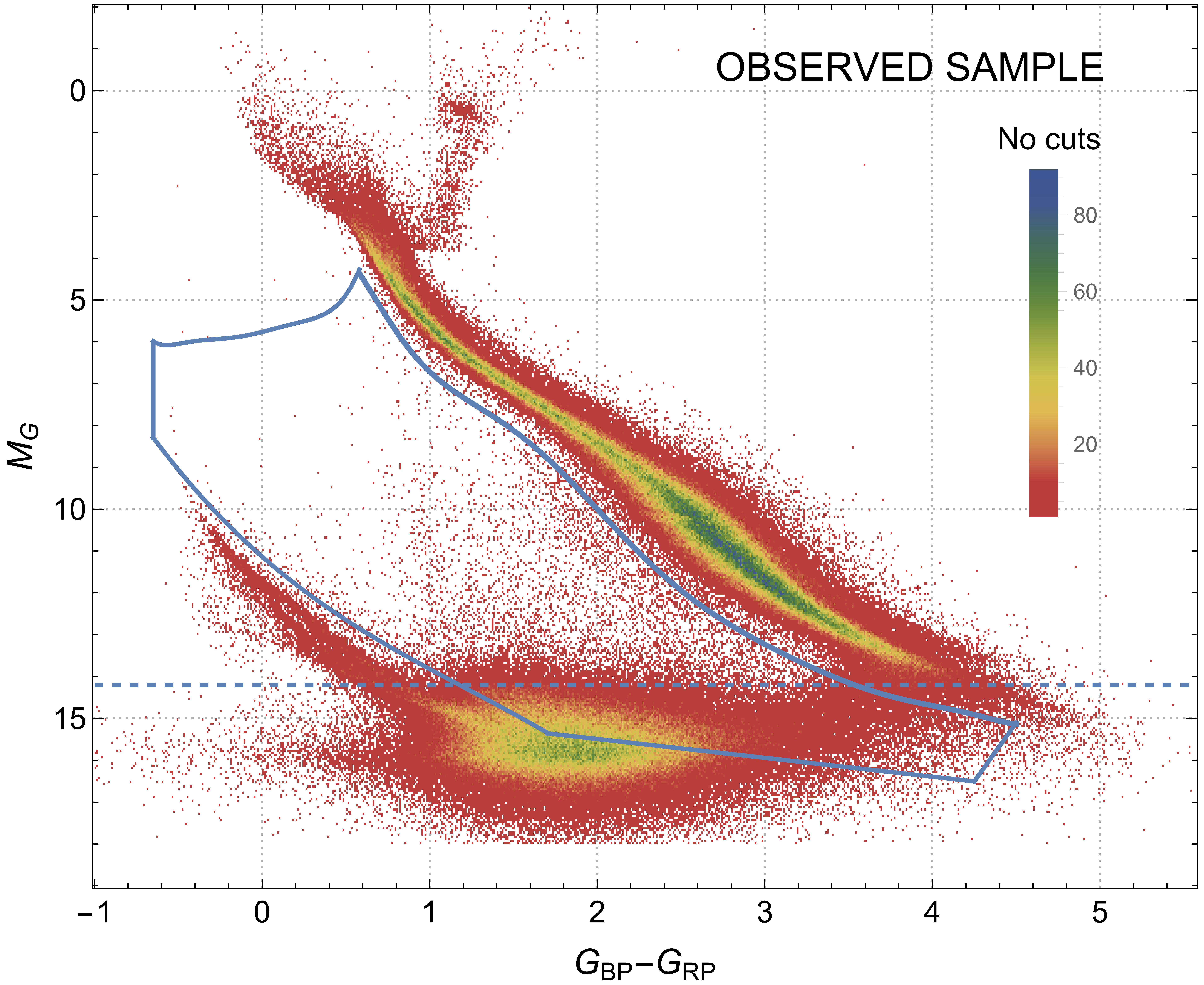}
\includegraphics[trim=0 0 0 0 clip=true, width=0.96\columnwidth]{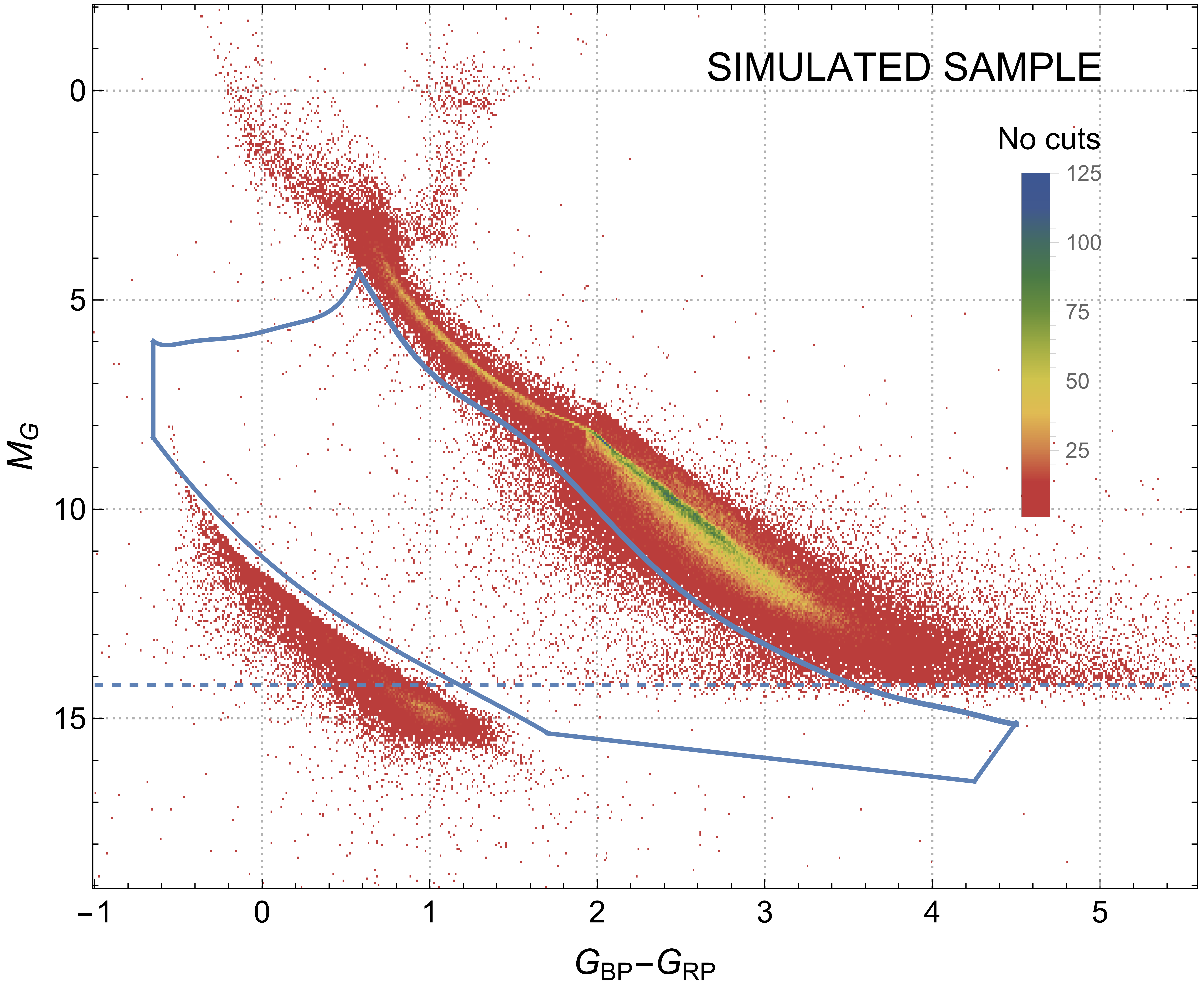}
\\
\includegraphics[trim=0 0 0 0 clip=true, width=0.96\columnwidth]{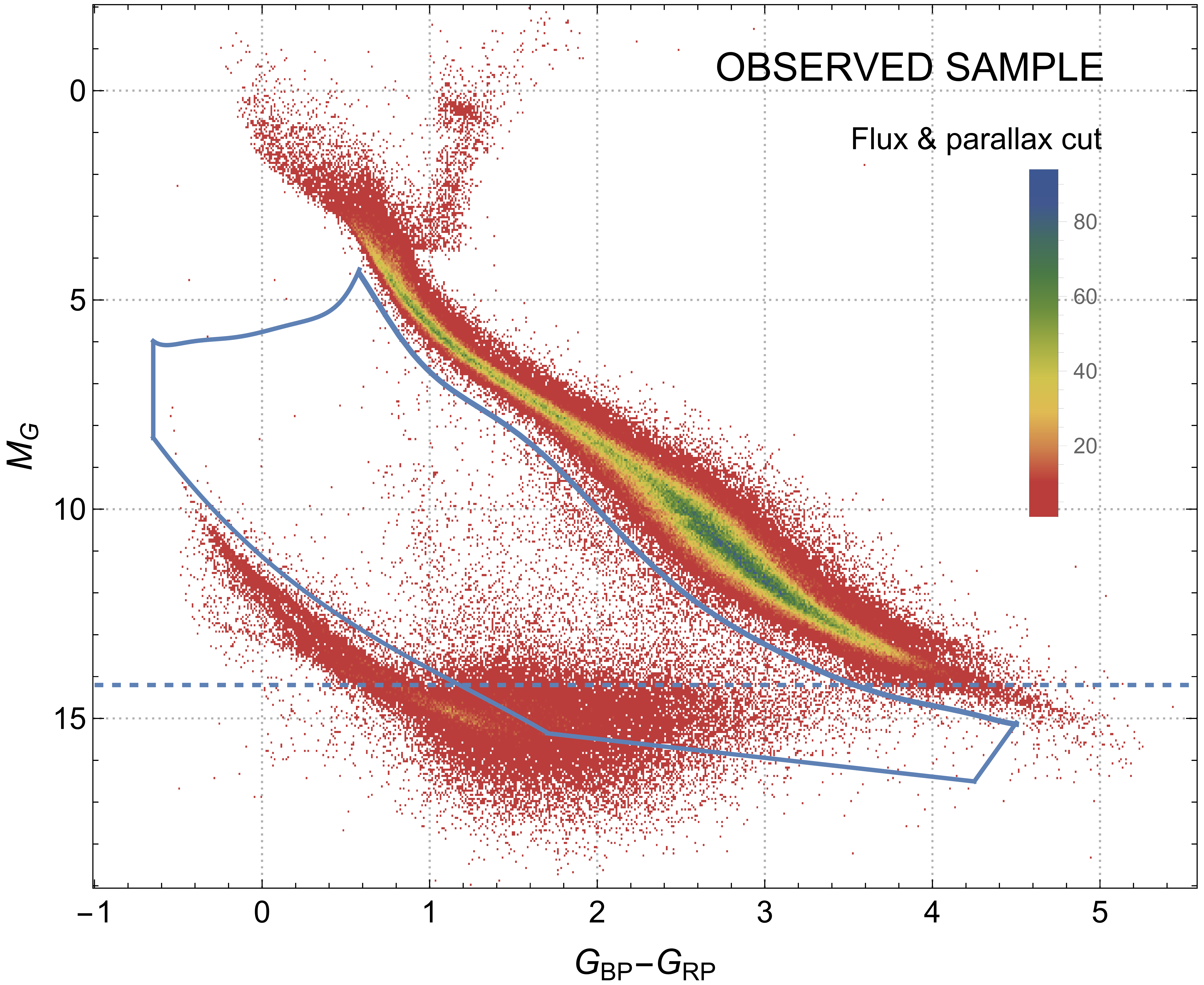}
\includegraphics[trim=0 0 0 0 clip=true, width=0.96\columnwidth]{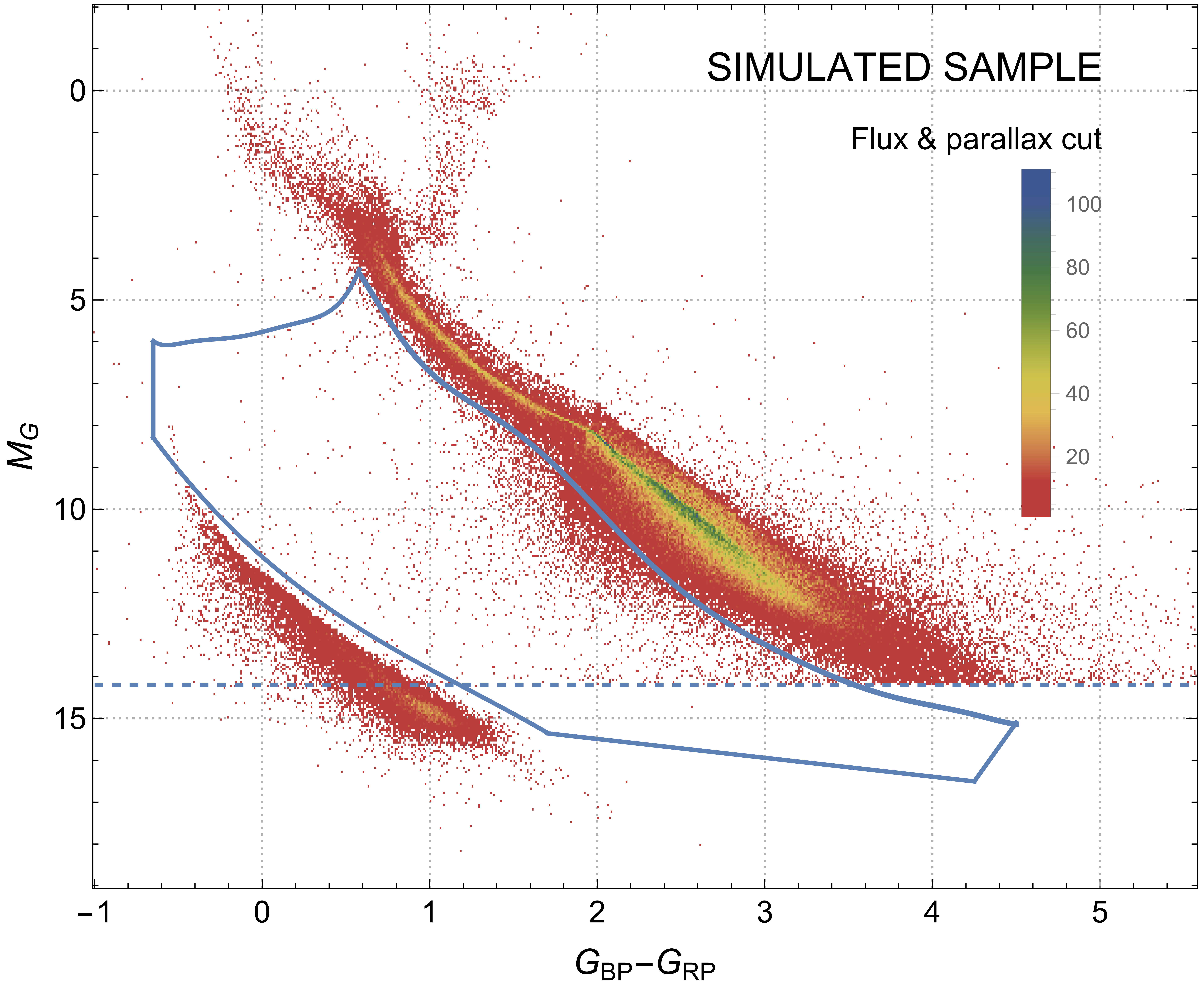}
\\
\includegraphics[trim=0 0 0 0 clip=true, width=0.96\columnwidth]{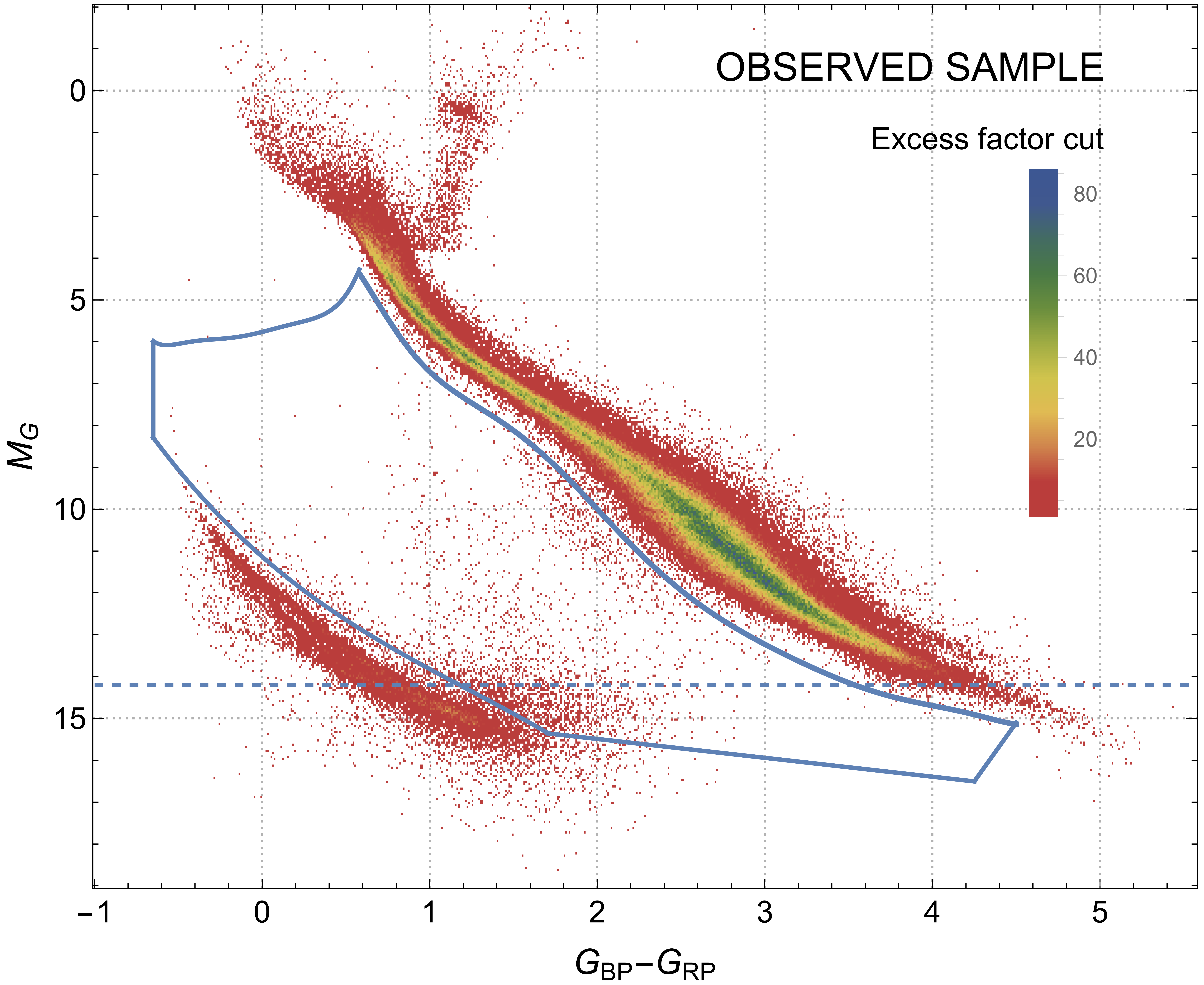}
\includegraphics[trim=0 0 0 0 clip=true, width=0.96\columnwidth]{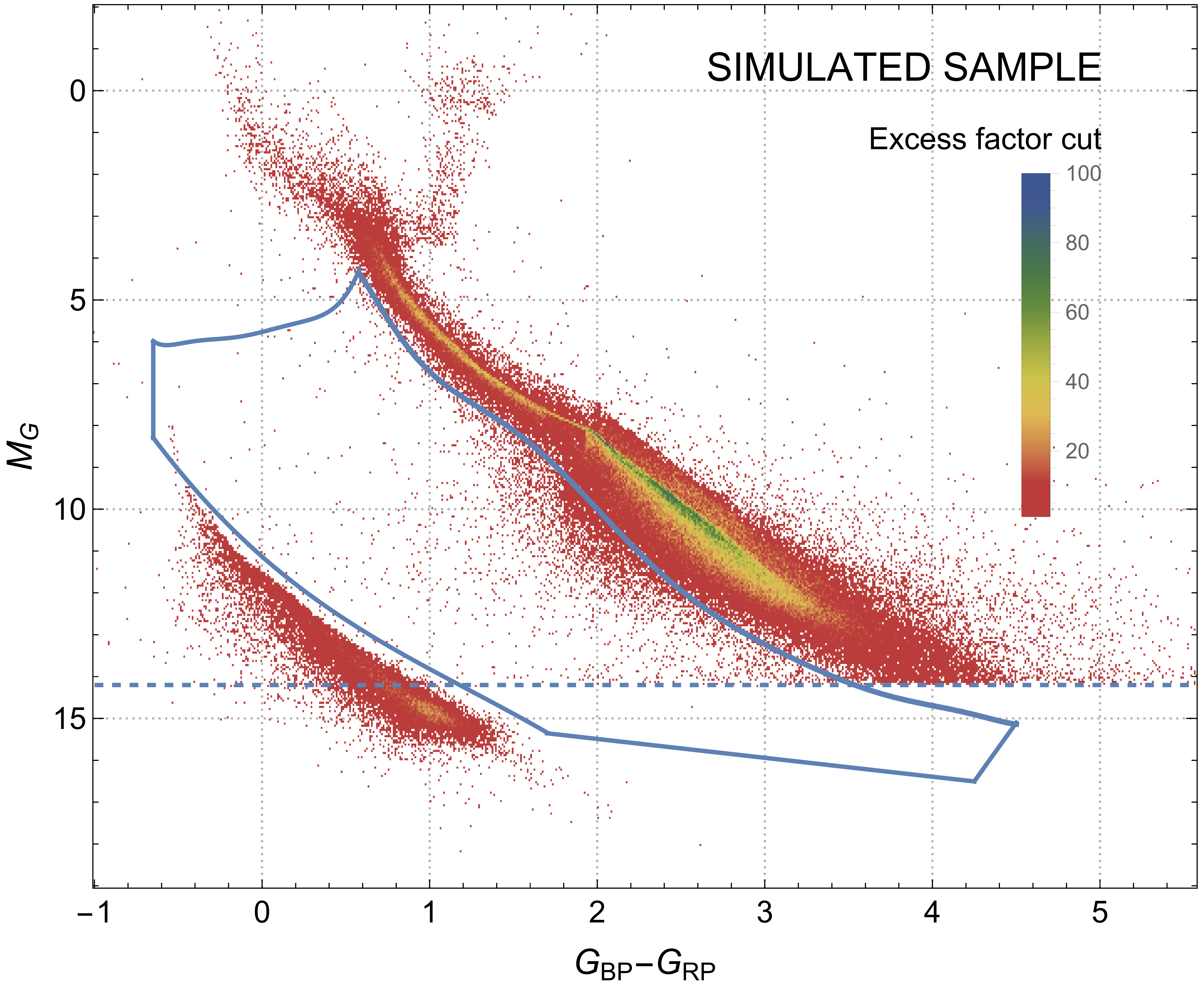}\\
\caption{Color-magnitude diagrams for the {\it Gaia} 100\,pc sample. The sample construction process is exemplified with color-magnitude diagrams for the observational sample (left panels) for the different selection phases (see Table \ref{t:tab1}): no cuts samples (upper panels), flux and parallax selection (middle panels), excess factor removal (bottom panels), and final WDMS sample (continued figure panels). Analogously, the corresponding diagrams of the simulated sample are shown (right panels). The WDMS region \citep{Rebassa+21} is marked with a solid blue line, while the dashed blue line depicts the magnitude limit of our simulations. Finally, the selection function (green line contours corresponding to 15\%, 36\% and 49\%) is derived from the observed sample from \citet[][bottom left panel, magenta dots; RM+21]{Rebassa+21}, and applied to a typical synthetic realization (bottom  right panel, magenta dots).
}
\label{f:Gaiavssim}
\end{figure*}

\begin{figure*}
\ContinuedFloat
\centering
\includegraphics[trim=0 0 0 0 clip=true, width=0.96\columnwidth]{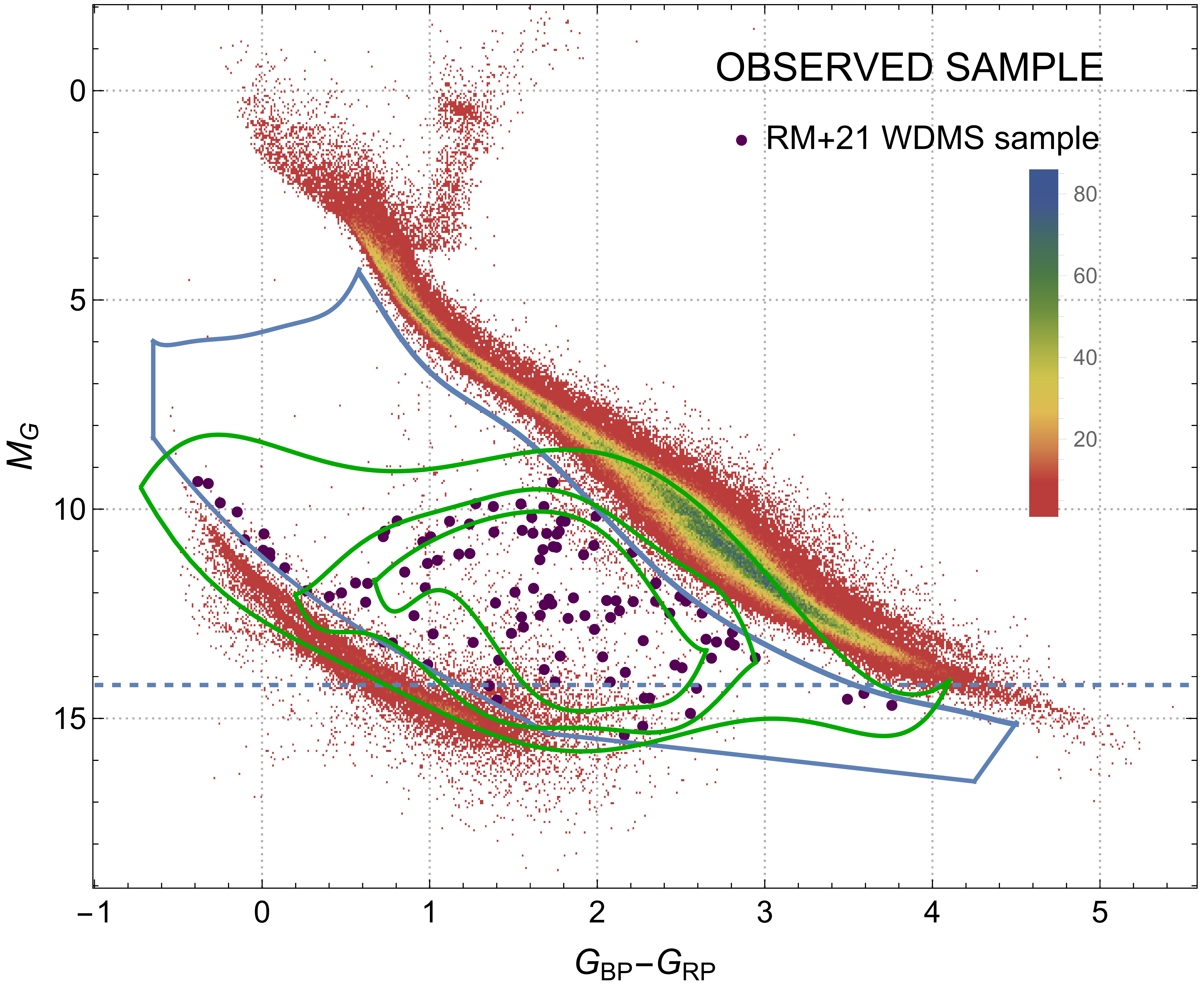}
\includegraphics[trim=0 0 0 0 clip=true, width=0.96\columnwidth]{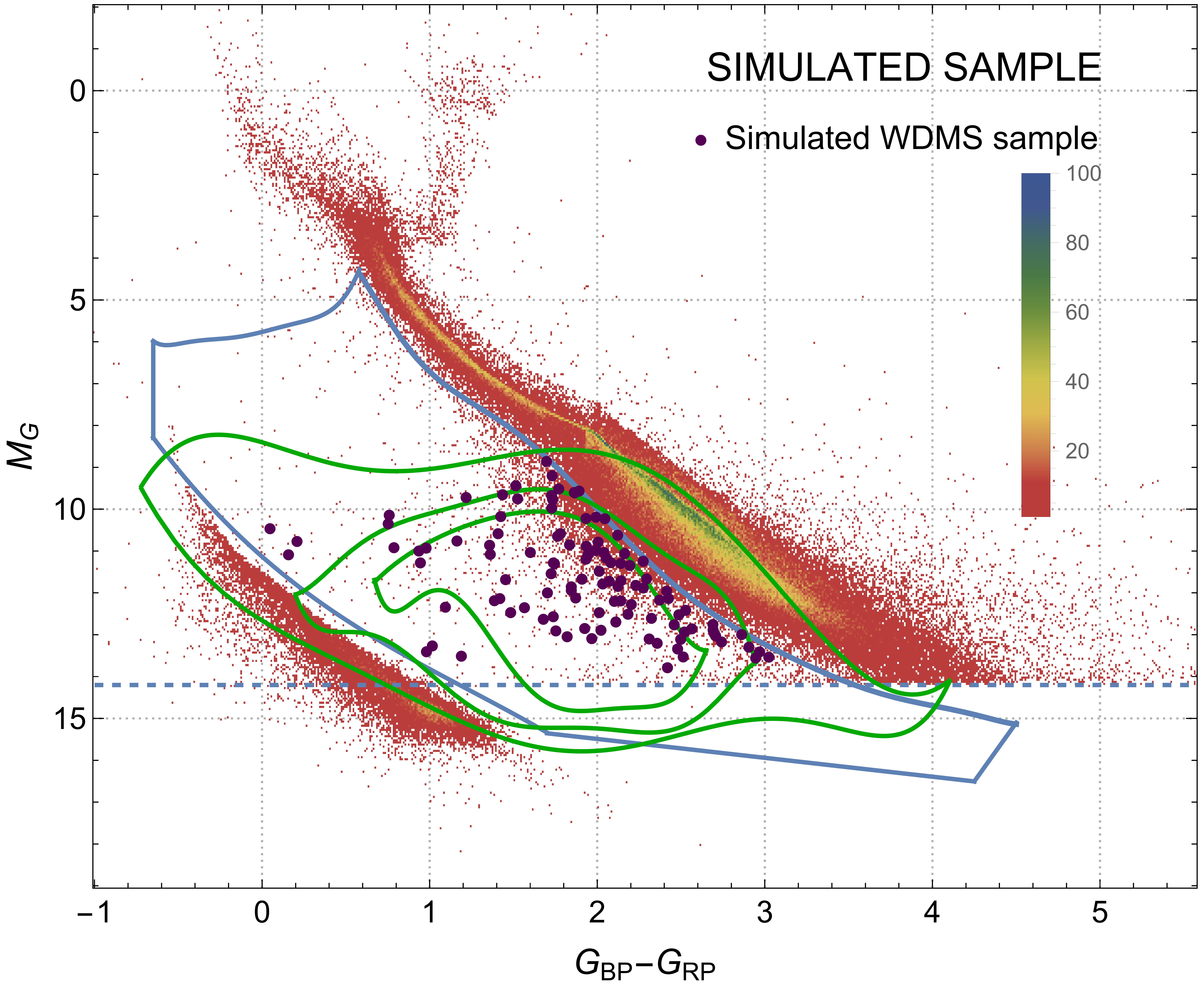}\\
\caption{continued.}
\end{figure*}

\begin{table*}[h]
\centering
\begin{tabular}{c|c|c}
\hline
\textbf{}         & \textbf{Observed sample} & \textbf{Simulated sample} \\ \hline \hline
No cuts           &   9839 ($\sim85000$)            & 7200 \\ \hline
Initial selection    & 7003 ($\sim20000$) & 5500\\ \hline
Excess factor cut & 1206 (2001)   & 4600 \\ \hline
WDMS pre-selection        & 213 (239)  & 240 \\ \hline
WDMS final selection       & 100 (112)  & 113  \\ \hline 
\end{tabular}
\caption{Number of observed and simulated objects for different selection criteria and magnitude $M_G<14.2\,$mag. The numbers in brackets in the left column include all objects with no restriction in $M_{G}$.}
\label{t:tab1}
\end{table*}

Following \emph{Gaia} performance, we distinguish between resolved and unresolved synthetic binaries. That is, a binary is considered to be unresolved if its angular separation is smaller than 2\arcsec \citep{Torres+22}, and thus being observed by \emph{Gaia} as just one stellar object instead of two separate well-characterized stars. It is worth noting that \emph{Gaia}'s resolution can nominally be even smaller \citep[$\sim0.5$\arcsec ][]{Gaia16}, especially for objects of equal brightness, which is not the general case for WDMS binary systems. Therefore, we have adopted a more conservative value to ensure completeness, as discussed in \citet{Torres+22}. For the purpose of studying these unresolved systems, the \texttt{MRBIN} code generates the angular separation for each system from the corresponding parallax and orbital separation, taking into account the eccentricity, inclination, and phase of the orbit. As a result some  systems may be classified as unresolved even though they have large periods (see Section \ref{ss:period}). For these synthetic unresolved systems it is required to consider new magnitude values that are calculated as an addition of the fluxes of both stars in the binary system. To do so, the absolute magnitudes previously obtained for each component are transformed into apparent magnitudes and subsequently converted into fluxes, which can then be added together and turned back into magnitudes.

In a final step, with the aim of closely mimicking the \emph{Gaia} observational data, we introduce both photometric and astrometric errors  following \citet{Riello+21}\footnote{\url{https://www.cosmos.esa.int/web/gaia/science-performance}} to the synthetic data previously generated. We also introduce a synthetic \emph{excess factor} \citep{Riello+21} to each target to quantify the expected flux excess arising from contamination from other sources. This excess is quantified from the observational data as $C^*=C-f(G_{\rm BP},G_{\rm RP})$, where $C=(I_{\rm BP}+I_{\rm RP})/I_G$ (\citealt{Evans+18}) and $f(G_{\rm BP},G_{\rm RP})$ is a function that gives the expected excess at a given color. Objects affected by no flux excess have a $C^*$ value close to zero. In our simulations, we use the previously obtained fluxes to determine the value of $C$ and $f(G_{\rm BP},G_{\rm RP})$ for each system/star, and all systems where $C^*< 1$ are considered unaffected by excess.

\subsection{Implementing the selection criteria and the selection function}
\label{ss:selection}

To ensure that our simulations apply the same selection criteria as those used for the observed \emph{Gaia} sample and to understand how these criteria impact our synthetic samples, we illustrate in Figure \ref{f:Gaiavssim}  how the most important cuts affect the observed and synthetic populations (left and right panels, respectively). These cuts include: the initial selection based on fluxes and parallaxes with relative errors under 10\% (top panels), the resulting sample after excluding objects with a high \emph{excess factor} parameter (middle panels) and the final sample after excluding single stars (bottom panels). In these plots, it is easy to distinguish between the population of single white dwarfs and main sequence stars. Between these two, there is a distinct region (following the selection criteria of \citealp[]{Rebassa+21}), where unresolved WDMS binaries are the predominant objects. We recall that our synthetic models are restricted to main sequence stars with $M_G$ absolute magnitudes above 14.2\,mag. Hence, for a proper comparison, only objects with $M_G$ absolute magnitudes below this value were considered. This cut is illustrated in Figure\,\ref{f:Gaiavssim} as a horizontal blue line.

In Table\,\ref{t:tab1} we provide the number of objects within the WDMS region that survive each selection cut for both the observed and simulated samples. When comparing these values %in Table\,\ref{t:tab1}
some differences arise. To begin with, the number of initially selected objects is considerably larger in the observed sample. This can be easily understood since observed data are associated to uncertainties that are challenging to simulate, such as wrong or spurious astrometric solutions, large values of the \emph{excess factor} parameter due to chance alignments, etc. Additionally, \emph{Gaia} observations reveal a noticeable cluster of objects in the lower intermediate region between white dwarfs and main sequence stars (see the top left panel of Figure \ref{f:Gaiavssim}). These objects are primarily located in the Galactic plane, therefore affected by crowding, and our current simulation code is unable to replicate them accurately. All these issues result in the inclusion of non WDMS binaries in our studied region, which are not represented in our simulations. Indeed, when the \emph{excess factor} cut is applied, the number of objects that are excluded is significantly larger for the observed sample ($\sim$83\%) than for the simulated sample ($\sim$16\%). It should be noted that the \emph{Gaia} observed sample contains many contaminated sources in the intermediate region, resulting in high color excesses or large astrometric errors. Our simulations, limited to Gaussian-distributed magnitude uncertainties, already remove most systems with large errors via the flux cut. Thus, applying the excess factor cut has a minimal effect, as few systems remain to be removed. However, after excluding single star contamination, the WDMS binary selection is very similar for both samples (of the order of 240 objects). These samples are further refined by taking into account the following aspects that affect the observed sample:  lack of sufficient photometry for performing the VOSA fits, detected UV excess due to magnetic activity, contamination from nearby sources and bad SED fitting (Section\,\ref{s:gaia}). The final observed sample is, thus, reduced to 112 candidates (100 if we consider only those with $M_G$<14.2\,mag). Implementing all these effects, which vary in nature and specific cases, in detail in our simulator would be a complex and overwhelming task. As a practical approach, to take them into account we introduce a selection function \citep{Rix2021}, $S_{\mathcal{C}}$, which assigns a probability to each member of our synthetic WDMS sample \(\mathcal{C}\) of belonging to the final sample. The selection function is built as a function of the luminosity and color, that is, $S_{\mathcal{C}}=S_{\mathcal{C}}(M_G, G_{\rm BP} - G_{\rm RP})$, proportionally to the density of WDMS observed candidates (magenta dots in bottom-left panel of Fig. \ref{f:Gaiavssim}) in the {\it Gaia} color-magnitude diagram. In the bottom-right panel of Fig. \ref{f:Gaiavssim} we depict the contour-levels (green lines)  of our selection function $S_{\mathcal{C}}$. We observe a higher concentration of objects toward the lower edge of the main sequence region, as theoretically expected \citep[see, for instance, Fig. 6 from][]{Rebassa+21}, compared to the corresponding dots in the observed sample. This indicates that the selection bias is even stronger than implemented. However, the significance of this observation is minor, as our analysis focuses on the total number of objects. That is, from the 240 simulated WDMS, the selection function selects 113 objects, which is in very good agreement with the observed sample of 100 objects.

%\section{Analysis of the different sub-populations}
\section{Analysis of the unresolved WDMS binary population}
\label{s:pop}

The synthetic single and binary 
populations, for both main-sequence and white dwarf stars, obtained from the model
described in Section \ref{ss:MonteCarlo} show reasonably good agreement with the 100 pc observed sample from \emph{Gaia} (bottom-right and left panels, respectively, of Fig. \ref{f:Gaiavssim}; see also Fig.\,\ref{f:KS_gaiaVSsim} of Appendix\,\ref{a:KStests} for a detailed comparison of the cumulative distributions for the magnitude $M_G$ and the color $G_{\rm BP}-G_{\rm RP}$ under different selection criteria). Moreover, according to \citet{Rebassa+21}, 100 \emph{Gaia} objects with $M_G<14.2$\,mag can be classified as unresolved WDMS in the intermediate region between white dwarfs and main-sequence stars, in good agreement with our final synthetic sample of 113 objects. These facts support the validity of the assumptions implemented in our model. Consequently, we take advantage of our \texttt{MRBIN} code to analyze the general content of the white dwarf population, with a particular focus on the unresolved white dwarf binary population. An analysis of the different aspects is presented as follows.

\begin{figure}
{\includegraphics[trim=0 0 0 0 clip=true, width=0.9\columnwidth]{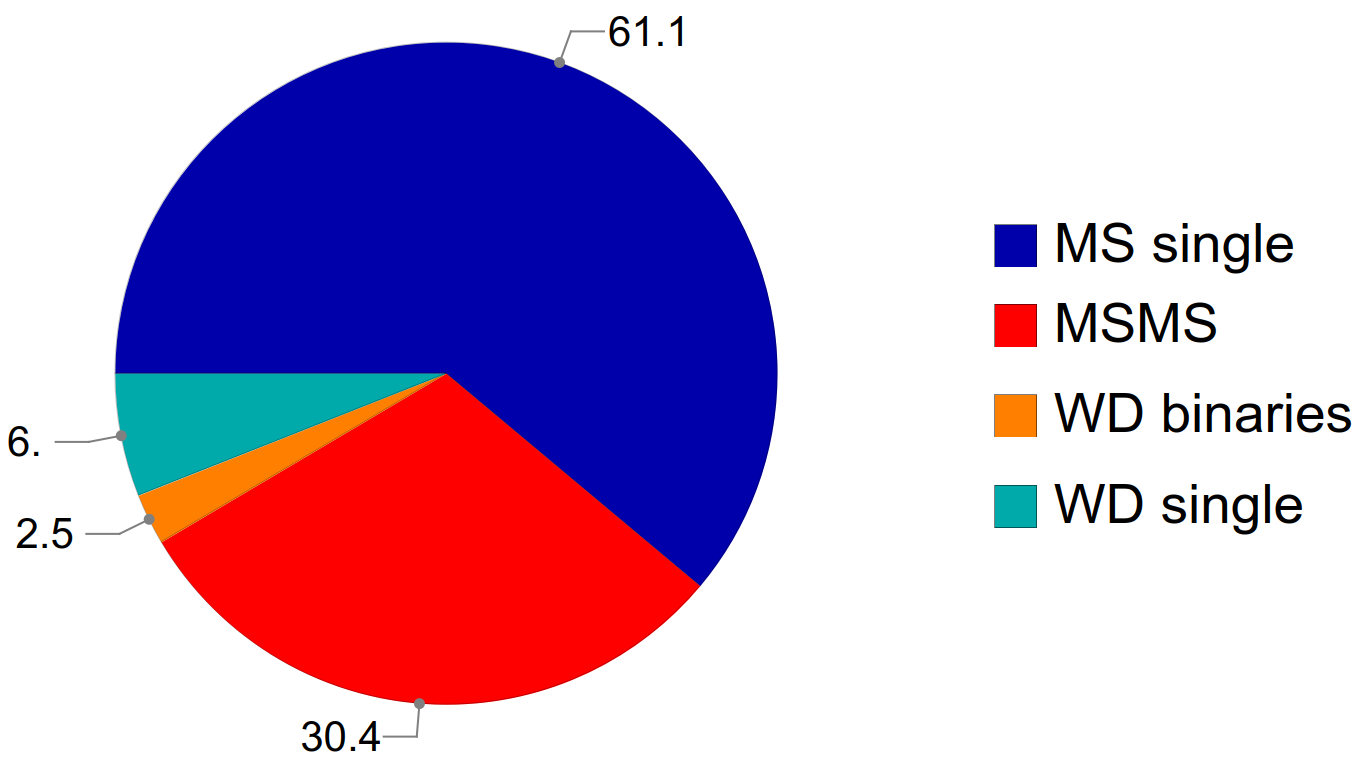}\\\includegraphics[trim=0 0 0 0 clip=true, width=1\columnwidth]{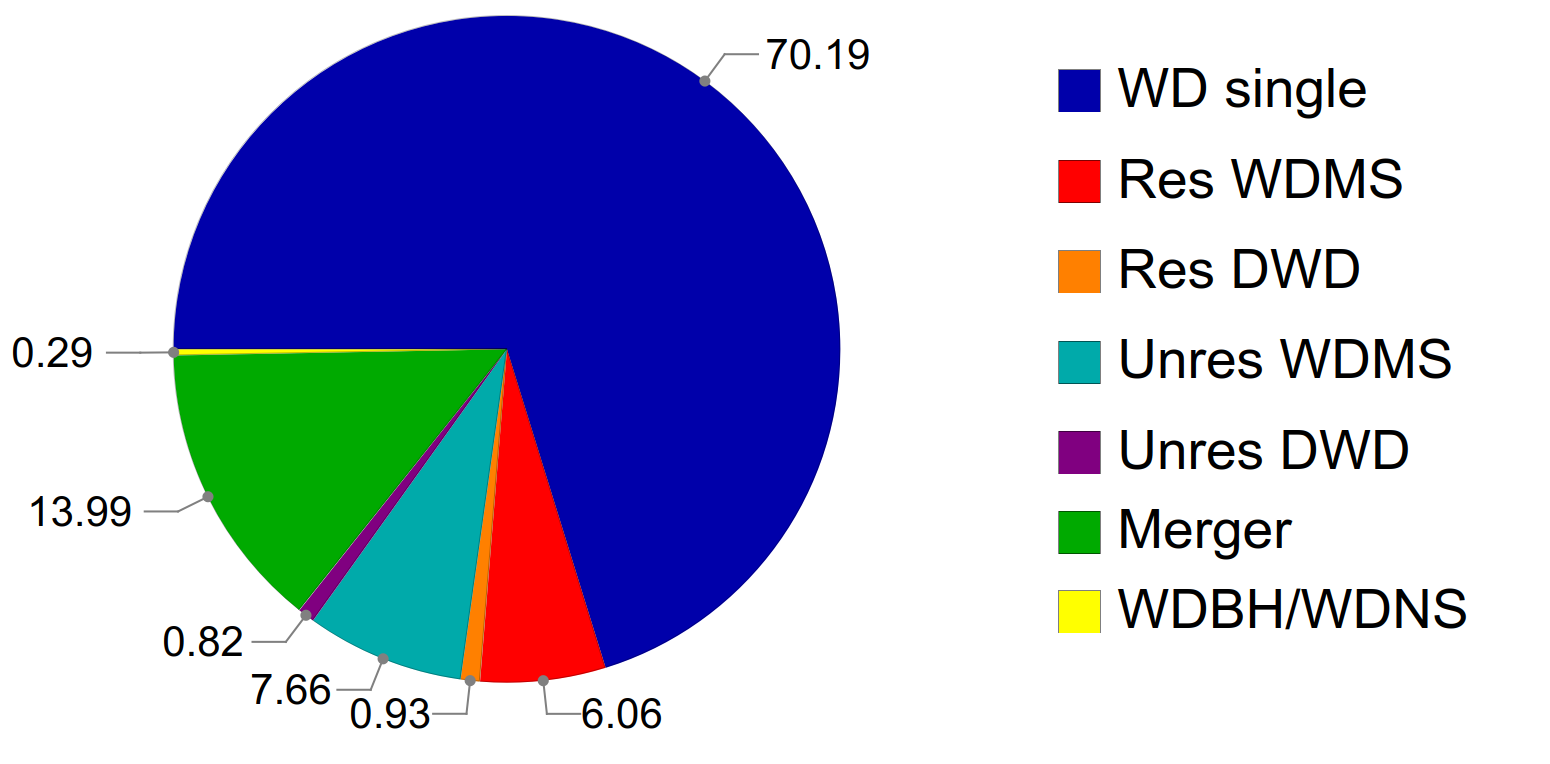}}
\caption{Pie chart showing the percentages of white dwarfs and main-sequence stars (top chart), and those of the white dwarf subpopulations (bottom chart) within 100 pc. The complete data can be found in Table \ref{tab:filter_pop} in Appendix \ref{a:fractions}.}
\label{f:pie}
\end{figure}

\subsection{General content of the white dwarf population}

The total number of stars observed at the current time in the {\it Gaia} color-magnitude diagram within 100\,pc in the white dwarf  region is 12\,718 \citep{Jimenez+23}. This value is used as a constraint for the normalization of our simulations. That is, our \texttt{MRBIN} code generates stars until the synthetic number of objects within the white dwarf region reaches this limit.

As previously stated, the input parameters used in our model reproduce the main features of the \emph{Gaia} color-magnitude diagram for the entire population of stars reasonably well. Moreover, the simulations are able to reproduce the constraints on the percentages of resolved WDMS and DWD in the white dwarf region \citep{Torres+22}, as well as the number of unresolved WDMS candidates in the intermediate region (Section \ref{ss:selection}). Hence, we are now able to analyze the contribution of the different sub-populations. 
We emphasize that, in this work, we consider only our reference model (that is, restricted to a specific set of parameters), and leave the analysis of possible dependencies with the model's input parameters for a future study. 

In Figure\,\ref{f:pie} we display a pie chart indicating the percentage of each individual population within the entire 100-pc {\it Gaia} color-magnitude diagram. In the top chart we show the contribution of the main populations: 67\% are single stars (61\% main-sequence and 6\% single white dwarfs), while the remaining 33\% are binary systems (30.5\% formed by two-main sequence stars, and 2.5\% containing a white dwarf). It is worth noting that, within the white dwarf population, 15.8\% of white dwarfs are in binary systems. Recalling that our normalization criteria is fixed to the number of total simulated objects in the white dwarf region, the previous percentages imply a total of $\sim100,000$ main-sequence stars generated within 100\,pc. This value is in agreement with \citet{Gaia+21}.

\begin{figure}
\centering
%%   \resizebox{\hsize}{!}
{\includegraphics[trim=0 0 0 0 clip=true, width=0.98\columnwidth]{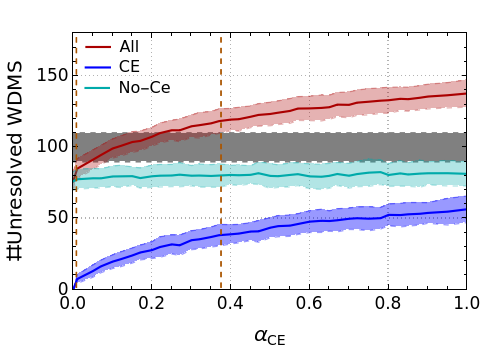}}
\caption{Number of unresolved WDMS systems as a function of the $\alpha_{\rm CE}$ parameter for binary systems in the WDMS region (red), showing also how many have undergone a common envelope episode (blue) and how many have not (green). Depicted as vertical orange dashed line are the lower and upper limits of $\alpha_{\rm CE}$ compatible with the observed value (gray). }
\label{f:alphanoce_final}
\end{figure}

In the bottom chart of Fig.\,\ref{f:pie} we analyze the resulting content of the different white dwarf sub-populations. As previously stated, most of them (84.2\%) are isolated white dwarfs, with 70.2\% formed as single white dwarfs and 14\% resulting from binary systems that have merged. This means that, of the total sample of isolated white dwarfs, around 17\% of them are products of merger episodes, in agreement with previous estimates  \citep[e.g.][]{Temmink+20}.

Regarding white dwarfs in unresolved binary systems, most are, as expected, WDMS systems, comprising 7.66\% of the total white dwarf population, while only a residual 0.82\% are DWD systems. This percentage of WDMS implies that nearly 1,000 unresolved systems are WDMS binaries. However, in the WDMS region, only 100 objects are ultimately identified. Thus, only a 10\% of this population is accessible by {\it Gaia} under the conditions imposed by \citet[][see Section\,\ref{s:gaia}]{Rebassa+21}. The remaining $\sim$90\% are mainly located in the main-sequence region of the {\it Gaia} color-magnitude diagram. These objects are challenging to be identified as WDMS, since the main sequence components outshine the white dwarfs in the optical. To identify such systems, it is common to use alternative methods involving ultraviolet photometry and/or spectroscopy \citep{Parsons+16, Rebassa+17, Ren+20, Anguiano+22, Nayak+24, Sidharth+24}.

The resolved WDMS and DWD populations will not be analyzed here since they are virtually identical to those of \citet{Torres+22} and were thus the dedicated subject of study in that work. Worth noting is that their percentages (6.06 for WDMSs and 0.93 for DWDs), as indicated in Figure\,\ref{f:pie}, are in very good agreement with the observed values in the white dwarf region  (6.31 for WDMSs and 1.18 for DWDs; \citealt{Torres+22}). 

Finally, the number of systems formed by a white dwarf and a black hole or a neutron star is less than a 0.3\%, a really low percentage as expected for a 100 pc volume limited sample.

\subsubsection{The common envelope treatment}
\label{ss:ce}

According to \citet{Rebassa+21}, 100 \emph{Gaia} objects with $M_G<14.2$\,mag can be classified as unresolved WDMS in the intermediate region between white dwarfs and main sequence stars. On the other hand, as we have shown (see Section \ref{ss:selection}), our reference model predicts a similar value of 113 objects. This value is actually the mean obtained from around 100 different Monte Carlo realizations, resulting a dispersion of $\pm 8$. Taking into account the dispersion derived from our simulation, and also the Poissonian error associated to the observed value, i.e., $100\pm10$ observed objects, we can conclude that our reference simulation is in agreement with the observed sample.

We recall that for our reference model we used a value  of the common envelope efficiency parameter of $\alpha_{\rm CE}=0.3$ \citep{Torres+22,Camacho+14}. However, depending on the adopted value of $\alpha_{\rm CE}$, we  can derive a different number of synthetic unresolved WDMS binaries.

To constrain the possible $\alpha_{\rm CE}$ values that yield synthetic unresolved WDMS samples in agreement with the observed value (taken as 100$\pm$10), 
we compare the total number of synthetic WDMS (after applying the selection function described in Section\,\ref{ss:selection}) to the observed one. In Figure\,\ref{f:alphanoce_final}, we depict the total number of synthetic WDMS across the full range of $\alpha_{\rm CE}$, with the average shown as a red line and the dispersion indicated by a red shaded area. Similarly, the observed value is shown in gray. As it can be seen, $\alpha_{\rm CE}$ values larger than 0.37 and lower than 0.02 yield synthetic samples that start to deviate from the observed number and, in principle, can be ruled out. This confirms the preference of low values for the common envelope efficiency during a common envelope episode, a result that has been found in the past by several authors \citep{Zorotovic+10, Toonen+13, Camacho+14, Cojocaru+17,  Iaconi+19, Zorotovic+22}.

Furthermore, our simulation also provides relevant information about how many unresolved WDMS systems that can be observed by {\it Gaia} within the WDMS region have gone through a common-envelope phase. In Fig.,\ref{f:alphanoce_final}, we show synthetic post common-envelope systems in blue and synthetic systems that evolved avoiding mass transfer episodes in green. As expected, the number of systems that do not evolve through a common envelope is practically the same, i.e. around 80 systems (initially 160 prior to the application of the selection function). However, the number of WDMS systems that passed through a common envelope phase increases with the adopted value of $\alpha_{\rm CE}$. This is a simple consequence of increasing the number of mergers during a common envelope event for lower $\alpha_{\rm CE}$ values, which imply a farther reduction of the orbit to eject the envelope.

We attempt to further constrain $\alpha_{\rm CE}$ by considering how many of the synthetic WDMS are eclipsing and comparing these numbers to the observed value. Of the 100 observed systems, Zwicky Transient Facility \citep[ZTF;][]{Decani+20} observations (constrained to a Dec limit of $-30^{\circ}$) indicate that 6 are totally eclipsing (\citealt{Brown+23}). By applying the same declination limit to our synthetic samples, we can determine how many of them are eclipsing by considering the minimum angle for total eclipse, $i_{\rm t}$, as

\begin{equation}
    i_{\rm t}=90^{\circ} - \arcsin\left( \frac{R_{\rm MS}-R_{\rm WD}}{a}\right),
\end{equation}
where $R_{\rm MS}$ and $R_{\rm WD}$ are the radius of the main-sequence and the white dwarf, respectively, and $a$ the semi-major axis of the binary system. If the inclination of a given binary as obtained from our code is larger than $i_t$, then the simulated WDMS can be considered as an eclipsing system. Notice also that this geometrical criterion promotes systems with small orbital separations (hence short periods, typically a few to tens of hours), thereby increasing the likelihood of eclipsing, as observed in such systems. There is a seventh system that is partially eclipsing according to ZTF data (Van Roestel, in preparation), which has not been considered in this work. It is also important to mention that the selection effects for identifying these systems are not yet well understood, and there no published estimates of their completeness. Consequently, we assume the six total eclipse systems as a lower limit for the 100 pc population.

\begin{figure}
\centering
%%   \resizebox{\hsize}{!}
{\includegraphics[trim=0 0 0 0 clip=true, width=0.98\columnwidth]{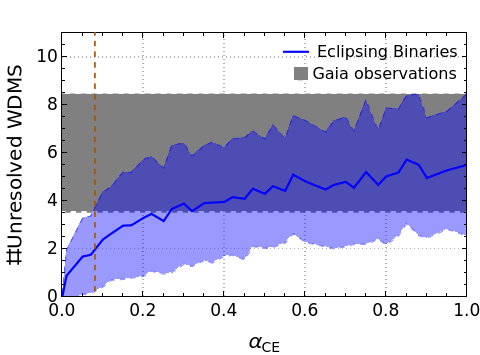}}
\caption{Number of eclipsing binaries as a function of the $\alpha_{\rm CE}$ parameter for binary systems  in the WDMS delimited region. As in Fig.\,\ref{f:alphanoce_final}, the vertical orange dashed line represents the lower limit of $\alpha_{\rm CE}$ compatible with the observed value (gray).}
\label{f:alphanoce_ec}
\end{figure}

The number of synthetic unresolved eclipsing WDMS binaries (blue line) as a function of $\alpha_{\rm CE}$ is illustrated in Figure\,\ref{f:alphanoce_ec}. As in Fig.\,\ref{f:alphanoce_final}, the shaded regions indicate the corresponding dispersions and the gray area corresponds to the observed value within the Poissonian error. Given the low number of eclipsing systems, the only constraint we can derive from Fig.\,\ref{f:alphanoce_ec} is that $\alpha_{\rm CE}$ should be higher than 0.08.

From the analysis performed in this section we thus constrain the $\alpha_{\rm CE}$ value to be in the range 0.1--0.4.

\begin{figure*}
\centering
{\includegraphics[trim=0 -20 0 0 clip=true, scale=0.58]{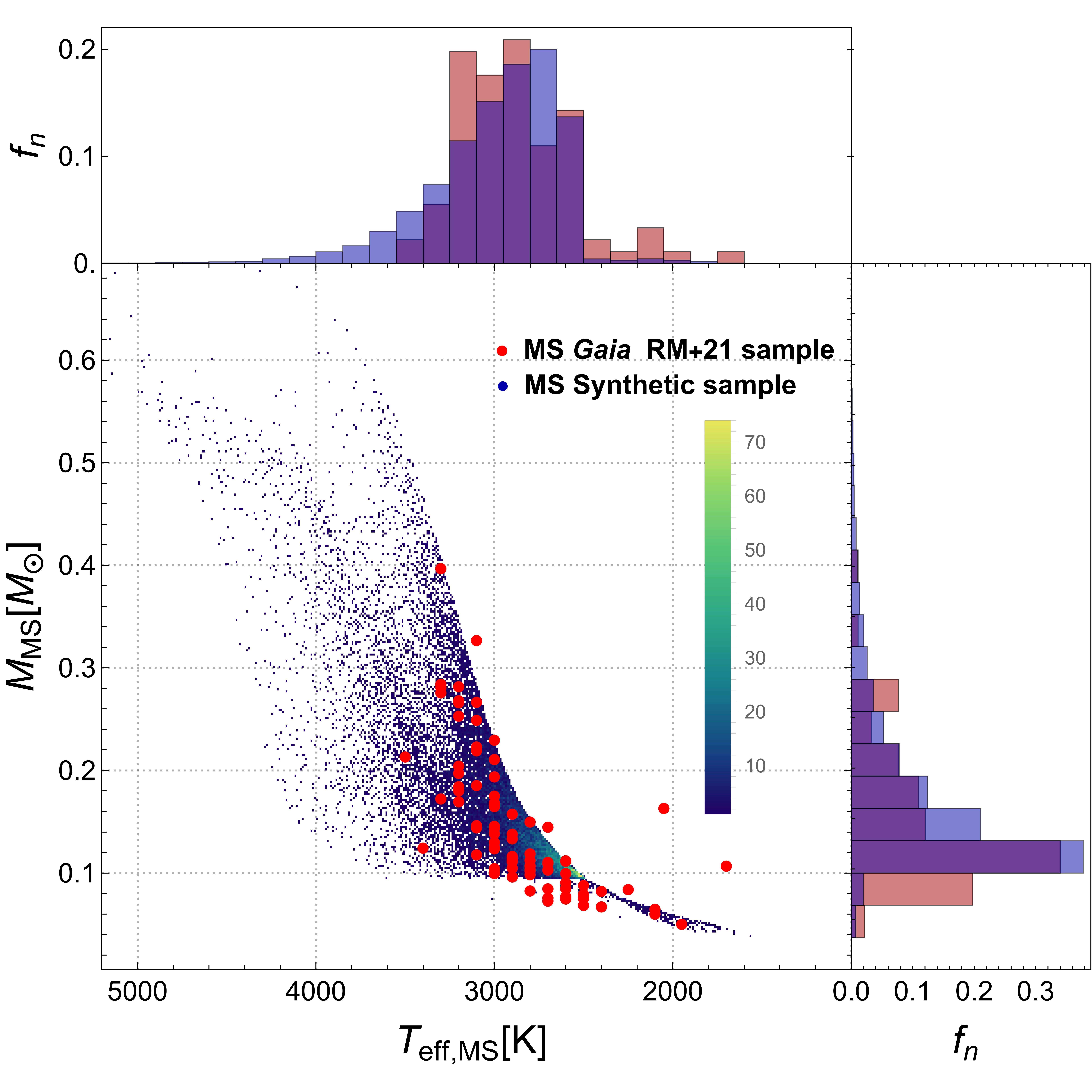}}
\includegraphics[trim=0 0 0 0 clip=true, scale=0.58]{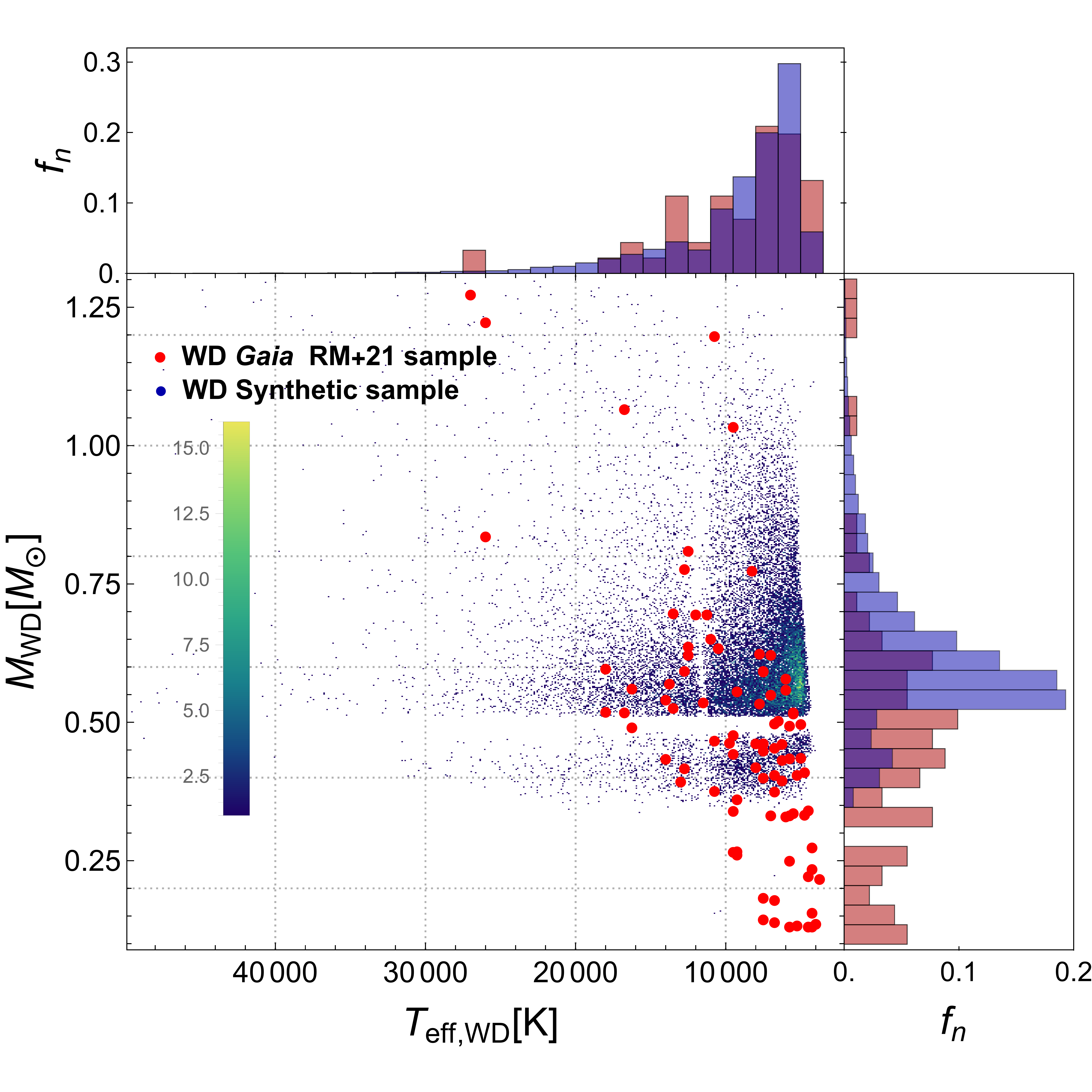}
\caption{Distribution of mass and $T_{\rm eff}$ for the main-sequence stars (top panel) and white dwarf component (bottom panel) of synthetic (blue points and histograms) and \emph{Gaia} observed (red points and histograms) WDMS systems.  For a better statistical comparison, 100 different synthetic realizations have been plotted.}
\label{f:WDTeff_M}
\end{figure*}

\subsubsection{Stellar parameter analysis}
\label{sss:ce}

In this Section, we aim to compare the stellar parameters, namely masses and effective temperatures, of the synthetic WDMS to those derived from the observed sample of \citet{Rebassa+21}. Based on the results from the previous Section, only synthetic models in which the $\alpha_{\rm CE}$ is approximately 0.1--0.4 should be used, and we fixed the value to $\alpha_{\rm CE}$=0.3 which corresponds to our reference model (note that no major differences were found when using other possible values). This  approach allows us to study the behavior of other parameters of our simulated sample.

Figure \ref{f:WDTeff_M} represents the mass-effective temperature plane of the simulated (blue dots) and observed (red dots) samples for the main sequence (top figure) and the white dwarf (bottom figure) components. For each figure, we plot the corresponding distribution of mass (right panels) and effective temperature (top panels) for both the observed and synthetic samples. Note that the simulated sample combines results from a large number (100) of realizations, providing a better characterization of the statistical properties of the synthetic sample.

Inspection of the top diagram of Fig. \ref{f:WDTeff_M} reveals a good agreement between the simulated and observed main sequence parameters, with the exception of a couple of observed objects (bottom-right region) that are clearly outliers. These are likely WDMS binaries with unreliable two-body fits from VOSA.  It is also worth noting that $\sim$10\% of the observed companions have mass estimates below 0.1\,M$_{\odot}$ and are probably brown dwarf candidates, objects that are not simulated by our code. The lack of such objects is also revealed when comparing the normalized distributions of mass and effective temperature (right and top panels). Additionally, we applied a Kolmogorov-Smirnov (KS) tests to the cumulative mass and effective temperature distributions (see top and bottom panels, respectively, of Figure \ref{f:KS_MS} of the Appendix \ref{a:KStests}) yield KS probabilities of 0.04 and 0.03, respectively. When comparing these values with a typical statistical significance level of $\alpha=0.05$, we can argue that the null hypothesis can be rejected, i.e., the distributions are not likely the same. However, given that the obtained p-values are close to the threshold, this fact suggests that there is evidence for a difference (as previously stated), but it is only moderate. If we perform the exercise of excluding both observed and synthetic objects below $0.1\,$M$_{\odot}$, the resulting KS probabilities are 0.40 for mass and 0.05 for effective temperature, indicating a strong agreement between the observed and simulated distributions.

Analogously, the observed and simulated white dwarf populations occupy similar regions in the mass-effective temperature plane (see bottom diagram of Fig. \ref{f:WDTeff_M}). Indeed, the normalized effective temperature distributions are rather similar, with a tendency of increasing objects for decreasing temperatures. The  KS probability obtained in this case is 0.17 (see the cumulative distributions in the bottom panel of Figure\,\ref{f:KS_WD} of the Appendix \ref{a:KStests}); clearly indicating compatibility between both distributions. However, the observed sample displays a clear overabundance of low-mass white dwarfs with respect to the synthetic sample.  
Indeed, the KS test applied to the cumulative mass distributions (top panel of Fig.\,\ref{f:KS_WD} of the Appendix \ref{a:KStests}) gives a null probability that  both samples being drawn from the same parent population. Even if we restrict our analysis to objects with masses above 0.35\,M$_\odot$, the probability remains null. Along with the possibility that other non-standard channels may exist for the formation of these objects, we speculate this is likely due to issues with the VOSA fits (Brown et al., in preparation, will provide a more detailed analysis), where discrepancies in white dwarf masses between different models could be as large as 20-30\%.

\begin{figure}
\centering
%%   \resizebox{\hsize}{!}
{\includegraphics[trim=0 0 0 0 clip=true, width=0.98\columnwidth]{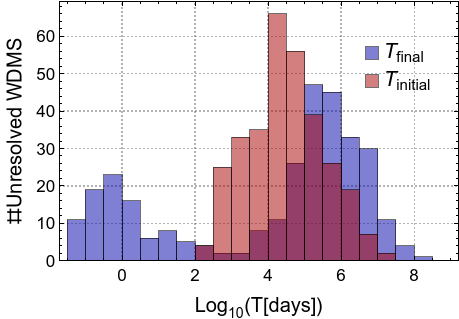}}
\caption{Initial (red) and final (blue) period distribution of simulated WDMS systems within the observed region.} 
\label{f:period}
\end{figure}

\begin{figure}
\centering
%%   \resizebox{\hsize}{!}
{\includegraphics[trim=0 0 0 0 clip=true, width=0.98\columnwidth]{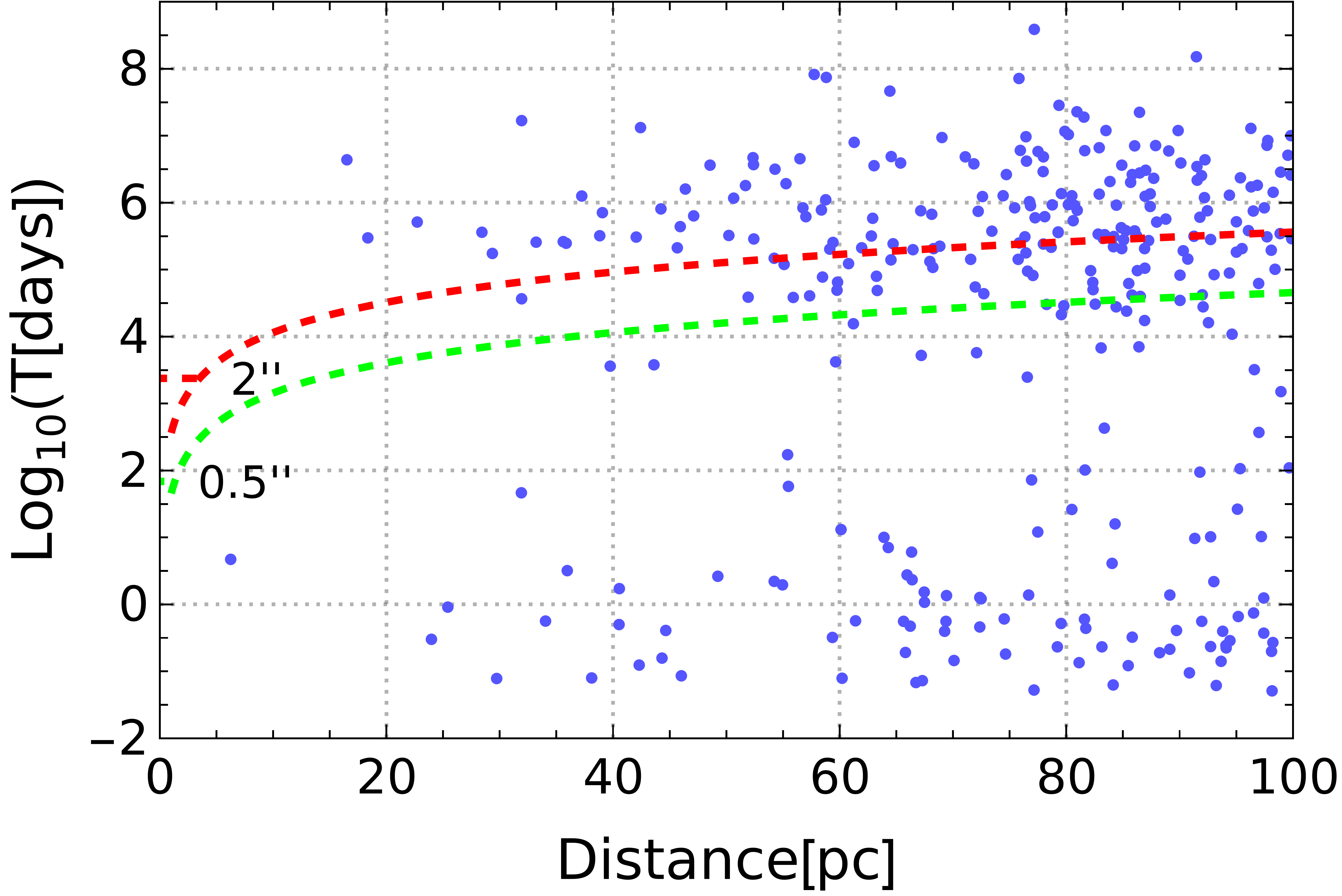}}
\caption{Distribution of the period versus the distance for the final simulated unresolved WDMS sample represented in Fig.  \ref{f:period}. Also plotted the nominal \emph{Gaia} angular resolution limit (0.5\arcsec; green dashed line) and the more conservative value adopted in our analysis (2\arcsec; red dashed line).} 
\label{f:distperiod}
\end{figure}

\subsection{Orbital period distribution and percentage of unresolved WDMS evolving through common envelope evolution}
\label{ss:period}

Although some differences arise between the simulated unresolved WDMS binary sample and the one observed by \emph{Gaia} and studied by \citet{Rebassa+21} -- see Section\,\ref{sss:ce}), we shall consider the synthetic sample as representative of the observed population. This allows us to estimate how many of the observed unresolved systems have evolved through a common envelope phase.

A clear way of illustrating this is by representing the orbital period distribution, which is shown in Figure \ref{f:period}. The initial main sequence binary orbital period distribution is illustrated in red and the resulting WDMS binary orbital period distribution in blue. It becomes clear that the final period distribution of WDMS binaries is bimodal, with those systems that evolved through common envelope having periods $\la$100 days and the non-interacting binaries having much longer periods. Note that in the latter case the orbital periods increase due to the mass loss of the white dwarf progenitors.  Furthermore, if we consider only those systems that have evolved through a common-envelope phase, we also observe a certain tendency to bimodality in their final distribution, consistent with recent analyses \citep{Ashley2019,Lagos2022}.

For a typical realization of our reference model, 35 WDMS binaries have evolved through common envelope evolution. Hence, our simulations predict that $\sim$30\% of the observed unresolved WDMS binary systems have gone through a common envelope episode during their evolution. Thus, we expect a similar percentage of observed WDMS to be post-common envelope binaries. Moreover, in Figure \ref{f:distperiod}, we plot the final period versus distance for a typical realization of our reference model. Even at closer distances, we can observe the existence of wide binary systems that remain unresolved. Specifically, for systems with long periods, $\log[T({\rm days})]>6$, resolution is particularly further limited by the inclination and eccentricity of the orbit. As a consequence, we can conclude that even for unresolved WDMS systems at close distances, it cannot be guaranteed that they are close binary systems.

\subsection{The unbiased unresolved population of single and binary stars}

The main purpose of this work has been to simulate the population of unresolved WDMS binaries accessible by \emph{Gaia} within 100 pc. To that end, we have restricted our simulated samples considering all possible observational biases affecting the population studied by \citet{Rebassa+21}, which decreased the original synthetic sample by $\sim$60\%. Hence, the considered observed WDMS binary sample is expected to be $\sim$40\% complete within the WDMS region. The results have shown that the synthetic population not only has nearly the same number of observed WDMS systems, but also that both samples have rather similar stellar parameters. This implies that the theoretical model adopted in our code for reproducing the population of Galactic single and binary stars can be considered as reliable. In consequence, we can argue that the unfiltered synthetic sample should be reliable of the underlying stellar population.

\begin{figure}
{\includegraphics[trim=0 0 0 0 clip=true, width=0.94\columnwidth]{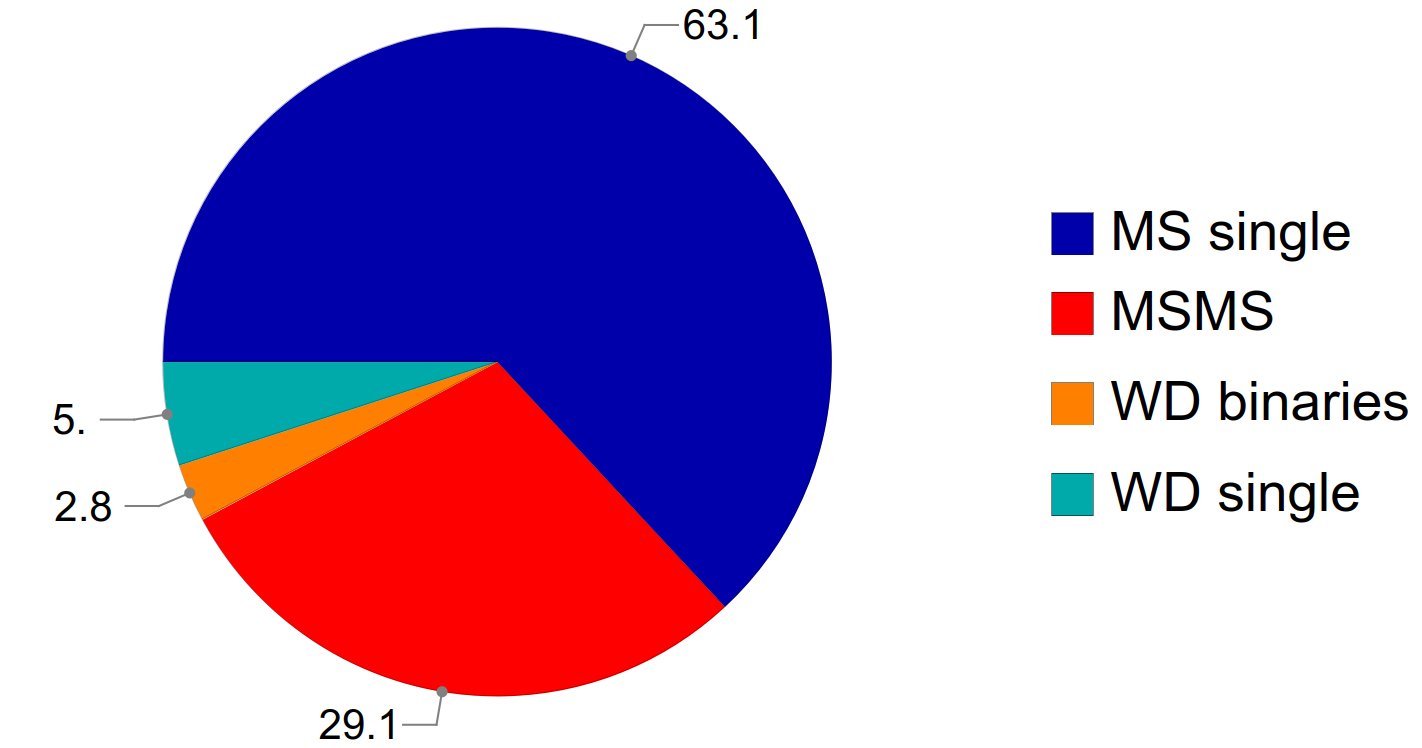}}\\\includegraphics[trim=0 0 0 0 clip=true, width=1\columnwidth]{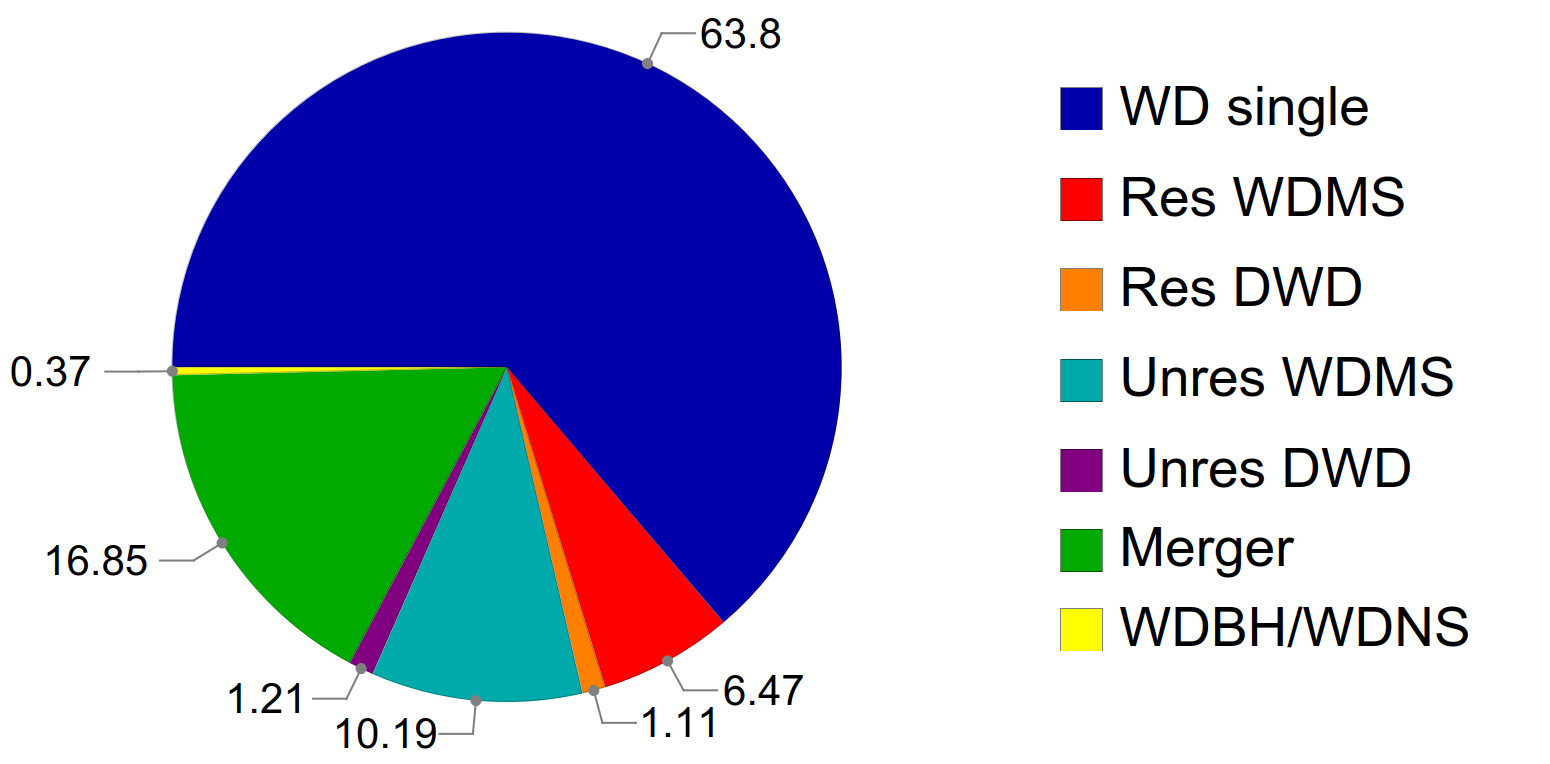}
\caption{As in Fig. \ref{f:pie} but for ideal conditions, that is, without considering any observational limits. The complete data can be found in Table \ref{tab:unfilter_pop} in Appendix \ref{a:fractions}.}
\label{f:pie2}
\end{figure}

Figure \ref{f:pie2} shows the percentages of single and binary stars (top panel) and the percentages of the different type of binaries (bottom panel) for the unfiltered synthetic population, that is,  assuming ideal conditions with no errors in flux, parallax or {\sl excess factor}, and without applying any final selection function. Interestingly, these percentages are rather similar to those derived from the restricted synthetic samples (see Figure \ref{f:pie}). 

\begin{figure*}[h!]
\centering
{\includegraphics[trim=12 0 0 0 clip=true, width=0.95\columnwidth]{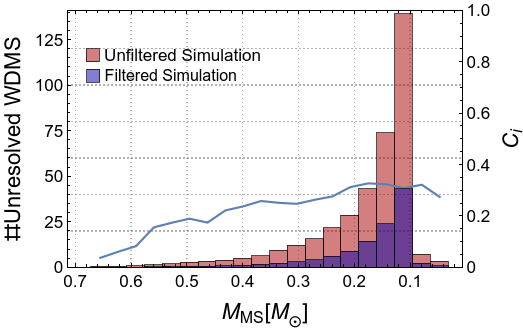}}
{\includegraphics[trim=-2 0 9 0 clip=true, width=0.96\columnwidth]{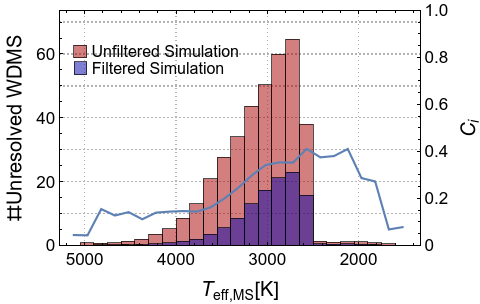}}
{\includegraphics[trim=6 0 0 0 clip=true, width=0.96\columnwidth]{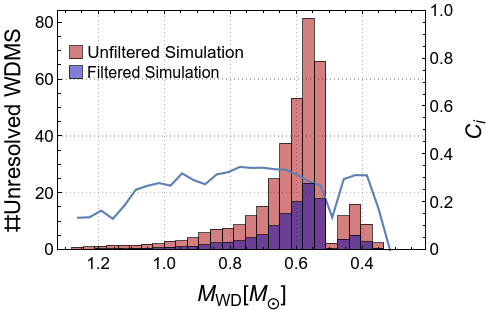}}
{\includegraphics[trim=6 0 10 0 clip=true, width=0.96\columnwidth]{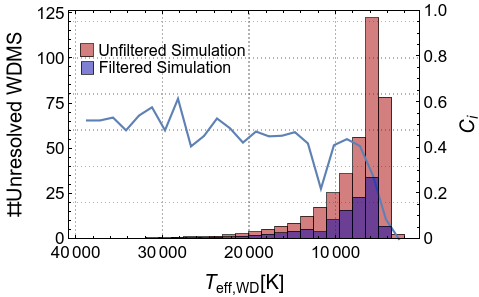}}
\caption{Mass and effective temperature distribution (left and right panels, respectively) of the main-sequence (top panels) and white dwarf (bottom panels) components of the unresolved synthetic systems, with (blue) and without (red) applying the observational filtering. The blue curves (right $Y-$axis) represent the completeness of each bin, $i$, in the distribution, obtained as $C_i=N_{\rm filtered}/N_{\rm unfiltered}$. Mean values from a total of 100 realization of our reference model were used.}
\label{f:unfil_mass_teff}
\end{figure*}

Regarding the unfiltered sample of unresolved WDMS binaries,  the total number of synthetic systems is 384$\pm$20 within the defined region of \citet{Rebassa+21}, which drops the completeness of the observed sample farther from $\sim$40\% to $\sim$26\%. Furthermore, the overall WDMS population consists of around 1,500 systems, representing only about $\sim 9\%$ of those accessible to {\it Gaia}. As previously stated, most of these objects are dominated by the flux of the main-sequence companion in the optical. Consequently, they are situated in the region of the color-magnitude diagram representative of single main-sequence stars, making them challenging to identify. A comparison between the stellar parameters of the synthetic WDMS with  and without  filtering the population is provided in Figure \ref{f:unfil_mass_teff} (blue and red histograms, respectively). Additionally, we estimate the completeness of each bin, $i$, as the ratio $C_i=N_{\rm filtered}/N_{\rm unfiltered}$. Although the mean completeness ($\sim 26\%$) is relatively low, the distribution of parameters derived from the {\it Gaia} sample can be considered representative of the WDMS population within that selection region.

\section{Conclusions}
\label{s:concl}

During the last two decades WDMS binaries have been proven to be excellent tools to improve our understanding of a wide variety of open issues in astronomy, ranging from the evolution of close compact binaries to the chemical and dynamical evolution of the Galactic disk. However, due to heavy observational selection effects, only a small percentage of WDMS are available for proper characterization and further analysis. On one hand, most WDMS binary samples are magnitude-limited. On the other hand, the main sequence components generally outshine the white dwarfs in the optical, which makes them difficult to be identified unless ultraviolet data is considered. This issue also affects recent volume-limited samples of unresolved WDMS systems from \emph{Gaia}.

In this work, we used a detailed Monte Carlo code, \texttt{MRBIN}, developed by our team to simulate the population of both single and binary main-sequence stars, as well as white dwarfs, in our solar neighborhood. Special emphasis has been placed on the analysis of the unresolved WDMS population, which has been filtered strictly following \emph{Gaia} observational selection criteria. The resulting synthetic samples have been compared with those obtained from the most homogeneous volume-limited catalog of WDMS binaries from \emph{Gaia} \citep{Rebassa+21}. This has allowed us to test the models used in the simulations and to evaluate the level of completeness of the observed sample.

Taking as a starting point the physical inputs obtained in \cite{Torres+22} in the fitting of the {\it Gaia} resolved WDMS and DWD populations, our synthetic sample aligns well with the distribution of observed unresolved WDMS in the \emph{Gaia} color-magnitude diagram. This fact reinforces the validity of our modeling approach, which provides a comprehensive description of the different subpopulations of our Solar neighborhood. In this sense, it is also important to emphasize that our simulator yields synthetic samples of single main sequence stars and white dwarfs, as well as resolved and unresolved WDMS binaries and DWDs whose overall numbers and fractions perfectly match the observed values. However, some discrepancies emerged when comparing the stellar parameters between the observed and synthetic unresolved WDMS, notably regarding the mass distribution of white dwarfs. We find that the synthetic population under-represents the lower-mass end of the observed distribution ($<0.5\,$M$_{\odot}$). This is likely related to uncertainties in the mass determination of these specific systems belonging to the observed sample.

By considering the synthetic WDMS sample without applying any observational cuts, we found that \emph{Gaia}'s observational constraints limit the detectable fraction of unresolved WDMS binaries to around 25\% in the intermediate region between white dwarfs and main-sequence stars in the \emph{Gaia} color-magnitude diagram. As mentioned before, this is not unexpected and clearly reveals that the majority of unresolved WDMS ($\sim$75\%) are located in the main sequence locus of the \emph{Gaia} color-magnitude diagram.

Our comparison between the observed and synthetic WDMS samples also places useful constraints on the common envelope efficiency parameter ($\alpha_{\rm CE}$), with best-fit values between 0.1 and 0.4. We find an overall agreement with observations for a value of $\alpha_{\rm CE}$=0.3, which confirms previous studies suggesting low values for $\alpha_{\rm CE}$. Additionally, our analysis indicates that $\sim$30\% of the observed {\it Gaia} unresolved WDMS population have undergone a common envelope episode.

This research strengthens the predictive power of population synthesis in stellar astrophysics and provides a foundation for interpreting future {\it Gaia} data releases and other white dwarf binary surveys.

\begin{acknowledgements}

We acknowledge the positive and constructive suggestions and comments of our anounymous referre. This work was partially supported by the MINECO grant  PID2020-117252GB-I00 and the PhD grant PRE2021-100503 funded by MICIU/AEI/10.13039/501100011033 and ESF+. This work has made use of data from the European Space Agency (ESA) mission {\it Gaia} (\url{https://www.cosmos.esa.int/gaia}), processed by the {\it Gaia} Data Processing and Analysis Consortium (DPAC, \url{https://www.cosmos.esa.int/web/gaia/dpac/consortium}). Funding for the DPAC has been provided by national institutions, in particular the institutions participating in the {\it Gaia} Multilateral Agreement.
\end{acknowledgements}
%%%%%%%%%%%%%%%%%%%%%%%%%%%%%%%%%%%%%%%%%%%%%%%%%%
\section*{Data Availability Statement}
The data underlying this article are available in the article.  Supplementary material will be shared on reasonable request to the corresponding author.


\bibliographystyle{aa}
\bibliography{PSBP}

\appendix
\clearpage

\section{Stellar population fractions within 100\,pc}
\label{a:fractions}

In Tables \ref{tab:filter_pop} and \ref{tab:unfilter_pop} we present the number of objects and the percentages (shown in brackets) of the different subpopulation of our simulation of the 100\,pc sample. Results correspond to a typical realization of our simulator, shown for cases with {\it Gaia} selection effects applied (Table \ref{tab:filter_pop}) and without them (Table \ref{tab:unfilter_pop}). Physical input parameters are those from \cite{Torres+22} and described in Section \ref{s:model}. The three last columns correspond to the total population (i.e. the complete color-magnitude diagram), the WDMS region (as defined in \cite{Rebassa2021b}), and the white dwarf region (as delimited in \cite{Jimenez+23}).

\begin{table*}[ht]
\centering
\caption{Synthetic filtered population}
\label{tab:filter_pop}
\begin{tabular}{ll|r|r|r}

\hline
            &  & \multicolumn{1}{|r|}{Total \#} & \multicolumn{1}{|r|}{WDMS region} & \multicolumn{1}{|r}{WD region} \\
\hline\hline
           \textbf{BINARIES}                                                            \\

\hline
               MSMS &        RESOLVED&    9608 (6.69) &    337 (7.36)   &   0 (0.00) \\
                &          UNRESOLVED&   34063 (23.71)&    987 (21.57)  &   0 (0.00) \\
          WDMS/WDRG &        RESOLVED&     739 (0.51) &     33 (0.72)   & 739 (6.57)  \\
           &               UNRESOLVED&    935 (0.65)  &    113 (2.47)   &   5 (0.04)  \\

                DWD &        RESOLVED&     115 (0.08) &      0 (0.00)   & 115 (1.02) \\
                 &         UNRESOLVED&     100  (0.08)&     11 (0.24)   &  89 (0.79)\\
          WDBH/WDNS &                &     42  (0.03) &      0 (0.00)   &  42 (0.37) \\
\hline\hline
            \textbf{SINGLES}                                            \\
\hline
                 MS &                &   86841 (60.46) &  3081 (67.34)  &   0 (0.00) \\
                 RG &                &     918 (0.64)  &     0 (0.00)   &   0 (0.00) \\  
                 WD &                &    8562 (5.96)  &     6  (0.13)  &8556 (76.08)\\
         WD MERGERS &                &   1707  (1.19) &      7 (0.15)   &1700 (15.10) \\

\hline
\\
\hline
   \multicolumn{1}{|r}{\textbf{TOTAL OBJECTS}} &                       & \multicolumn{1}{r|}{\textbf{143632}} &              \multicolumn{1}{r|}{\textbf{4575}}          &       \multicolumn{1}{r|}{\textbf{11246}}                   \\
\hline
\end{tabular}
\end{table*}

\begin{table*}[ht]
\centering
\caption{Synthetic unfiltered population}
\label{tab:unfilter_pop}
\begin{tabular}{ll|r|r|r}

\hline
            &  & \multicolumn{1}{|r|}{Total \#} & \multicolumn{1}{|r|}{WDMS region} & \multicolumn{1}{|r}{WD region} \\
\hline
\hline
           \textbf{BINARIES}                                                            \\
           \hline
               MSMS &    RESOLVED   &   10388 (5.47) &   367 (5.78) &    0 (0.00)  \\
                &       UNRESOLVED  &   44778 (23.59)&  1305 (20.55) &   0 (0.00) \\
          WDMS/WDRG &    RESOLVED   &     966 (0.51) &    42 (0.66) &  966 (7.22) \\
           &            UNRESOLVED  &    1522 (0.80) &   384 (6.05) &    8 (0.06) \\
                DWD &    RESOLVED   &     166 (0.09) &     0 (0.00) &  166 (1.24) \\
                 &      UNRESOLVED  &     180 (0.09) &    17 (0.27) &  163 (1.22) \\
          WDBH/WDNS &               &      55 (0.03) &     0 (0.00) &   55 (0.40) \\
\hline\hline
            \textbf{SINGLES}                                              \\
\hline
                 MS &               &  118444 (62.41) & 4214 (66.37) &   0 (0.00)  \\
                 RG &               &    1251 (0.66) &    0 (0.00)  &    0 (0.00) \\  
                 WD &               &    9528 (5.02) &    9 (0.14)  & 9519 (71.11) \\
         WD MERGERS &               &    2517 (1.33) &   11 (0.17)  & 2506 (18.73) \\
                 \hline
\\
\hline
   \multicolumn{1}{|r}{\textbf{TOTAL OBJECTS}} &                       & \multicolumn{1}{r|}{\textbf{189795}} &              \multicolumn{1}{r|}{\textbf{6349}}          &       \multicolumn{1}{r|}{\textbf{13383}}                   \\
\hline
\end{tabular}
\end{table*}

\clearpage

\section{Kolmogorov-Smirnov tests}
\label{a:KStests}

%\FloatBarrier

\begin{figure*}[h!]
\centering
\includegraphics[trim=0 0 0 0 clip=true, width=0.96\columnwidth]{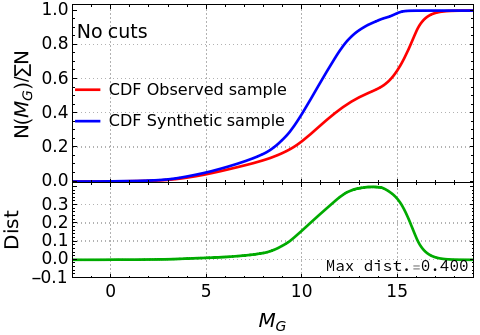}
\includegraphics[trim=0 0 0 0 clip=true, width=0.96\columnwidth]{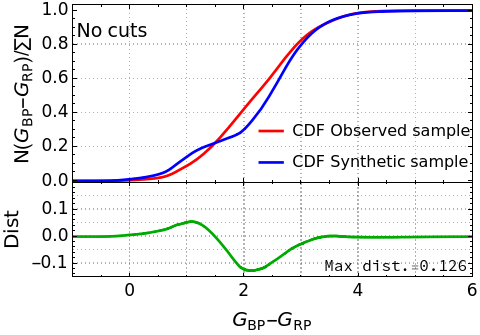}
\\
\includegraphics[trim=0 0 0 0 clip=true, width=0.96\columnwidth]{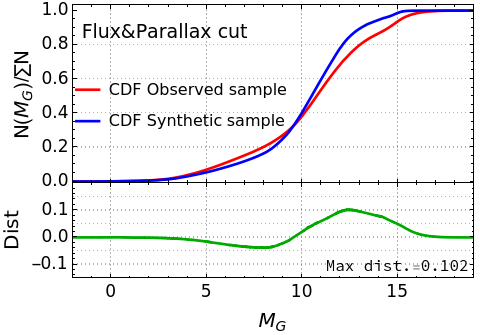}
\includegraphics[trim=0 0 0 0 clip=true, width=0.96\columnwidth]{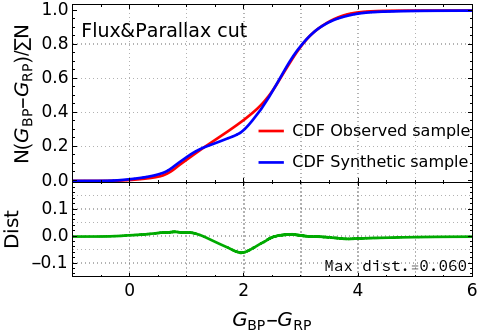}
\\
\includegraphics[trim=0 0 0 0 clip=true, width=0.96\columnwidth]{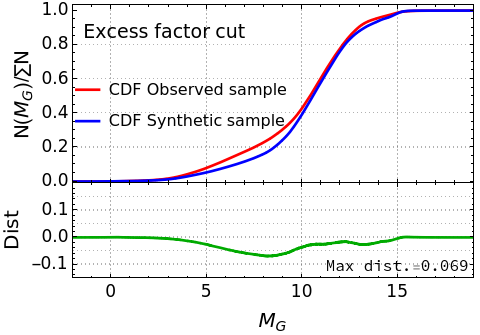}
\includegraphics[trim=0 0 0 0 clip=true, width=0.95\columnwidth]{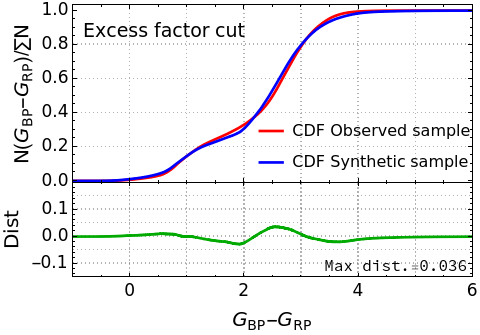}\\
\caption{Cumulative Distribution Function (CDF) of the synthetic sample (blue line) and the observed {\it Gaia}  sample from Fig. \ref{f:Gaiavssim} (red line) for the magnitude $M_G$ (left panels) and the color $G_{\rm BP}-G_{\rm RP}$ (right panels)  under different selection criteria.  For a more quantitative comparison, the distance (DIST; green line) between the CDFs is also plotted. As expected, the agreement between the observed and synthetic samples improves as the different selection criteria are applied.}
\label{f:KS_gaiaVSsim}
\end{figure*}

\begin{figure}[ht!]
 %  \resizebox{\hsize}{!}
{\includegraphics[trim=0 0 0 0 clip=true, width=1.0\columnwidth]{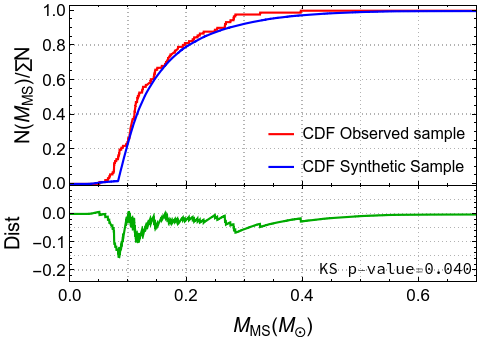}}
{\includegraphics[trim=0 0 0 0 clip=true, width=1.0\columnwidth]{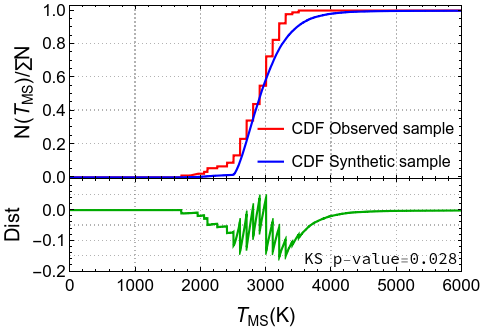}}
\caption{Kolmogorov-Smirnov tests for mass (top) and effective temperatures (bottom). In each panel, the top subpanels represents the Cumulative Distribution Function (CDF) of the main sequence component of the observed {\it Gaia} sample \citep[][red line]{Rebassa+21} and the synthetic sample (blue line), and the bottom subpanels display the distance (DIST; green line) between the CDFs}.
\label{f:KS_MS}
\end{figure}

\begin{figure}
%   \resizebox{\hsize}{!}
{\includegraphics[trim=0 0 0 0 clip=true, width=1.0\columnwidth]{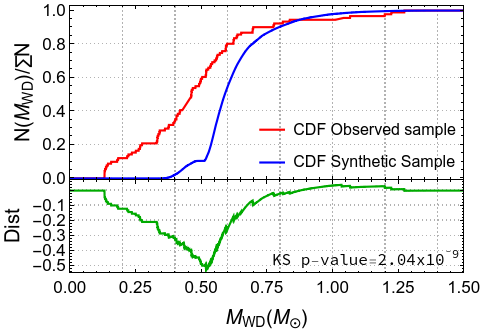}}
\includegraphics[trim=0 0 0 0 clip=true, width=1.0\columnwidth]{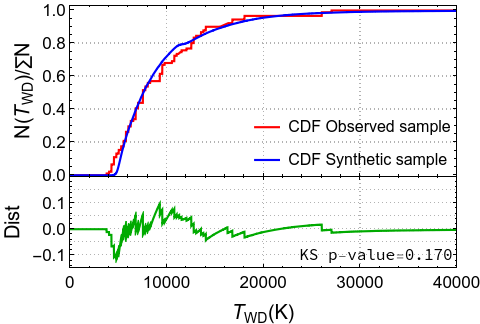}
\caption{As in Fig. \ref{f:KS_MS}, but for the white dwarf component.}
\label{f:KS_WD}
\end{figure}

%-------------------------------------------------------------------

\end{document}